\documentclass[twocolumn,epsfig,graphics,noshowpacs,floatfix,nofootinbib]{revtex4}

\usepackage{amsmath,amsfonts,amssymb,graphics,graphicx,epsfig,color,times,bbm}
\usepackage{amsthm}
\usepackage{psfrag}
\usepackage{braket}
\AtBeginDocument{\usepackage{booktabs}}
\graphicspath{{figures/}}
\usepackage{qcircuit}
\usepackage[hidelinks]{hyperref}
\hypersetup{colorlinks=false}
\usepackage{array}
\newcolumntype{P}[1]{>{\centering\arraybackslash}p{#1}}
\newcolumntype{M}[1]{>{\centering\arraybackslash}m{#1}}
\usepackage{graphicx}

\usepackage[framemethod=tikz]{mdframed}
\usepackage{hyperref}

\definecolor{warning_bgcol}{RGB}{252,248,229}
\definecolor{warning_textcol}{RGB}{111,89,54}
\definecolor{warning_linecol}{RGB}{252,248,229}
\definecolor{danger_bgcol}{RGB}{239,223,222}
\definecolor{danger_textcol}{RGB}{128,60,57}
\definecolor{danger_linecol}{RGB}{239,223,222}
\definecolor{success_bgcol}{RGB}{224,237,216}
\definecolor{success_textcol}{RGB}{68,104,60}
\definecolor{success_linecol}{RGB}{224,237,216}
\definecolor{info_bgcol}{RGB}{220,237,246}
\definecolor{info_textcol}{RGB}{58,100,126}
\definecolor{info_linecol}{RGB}{220,237,246}

\mdfdefinestyle{box_style}{
  skipabove=.7\baselineskip,
  skipbelow=.7\baselineskip,
  innertopmargin=.65\baselineskip,
  innerbottommargin=.65\baselineskip,
  innerleftmargin=.5\baselineskip,
  innerrightmargin=.5\baselineskip,
  splittopskip=1.5\baselineskip,
  splitbottomskip=\baselineskip,
  roundcorner=.3\baselineskip
}
\mdfdefinestyle{warning_style}{
  style=box_style,
  backgroundcolor=warning_bgcol,
  linecolor=warning_linecol,
  fontcolor=warning_textcol,
}
\mdfdefinestyle{success_style}{
  style=box_style,
  backgroundcolor=success_bgcol,
  linecolor=success_linecol,
  fontcolor=success_textcol,
}
\mdfdefinestyle{danger_style}{
  style=box_style,
  backgroundcolor=danger_bgcol,
  linecolor=danger_linecol,
  fontcolor=danger_textcol,
}
\mdfdefinestyle{info_style}{
  style=box_style,
  backgroundcolor=info_bgcol,
  linecolor=info_linecol,
  fontcolor=info_textcol,
}

\begin{document}

\title{Fusion-based quantum computation}

\author{Sara Bartolucci}
\author{Patrick Birchall}
\author{Hector Bomb\'in\footnote{these authors contributed equally} }
\author{Hugo Cable}
\author{Chris Dawson}
\author{Mercedes Gimeno-Segovia}
\author{Eric Johnston}
\author{Konrad Kieling}
\author{Naomi Nickerson \footnotemark[1]} \email{naomi@psiquantum.com}
\author{Mihir Pant \footnotemark[1]} \email{mihir@psiquantum.com}
\author{Fernando Pastawski}
\author{Terry Rudolph}
\author{Chris Sparrow}

\affiliation{PsiQuantum, Palo Alto}
\date\today

\begin{abstract}
We introduce fusion-based quantum computing (FBQC) - a model of universal quantum computation in which entangling measurements, called \textit{fusions}, are performed on the qubits of small constant-sized entangled \textit{resource states}. We introduce a stabilizer formalism for analyzing fault tolerance and computation in these schemes. This framework naturally captures the error structure that arises in certain physical systems for quantum computing, such as photonics.
FBQC can offer significant architectural simplifications, enabling hardware made up of many identical modules, requiring an extremely low depth of operations on each physical qubit and reducing classical processing requirements. We present two pedagogical examples of fault-tolerant schemes constructed in this framework and numerically evaluate their threshold under a hardware agnostic fusion error model including both erasure and Pauli error. We also study an error model of linear optical quantum computing with probabilistic fusion and photon loss. In FBQC the non-determinism of fusion is directly dealt with by the quantum error correction protocol, along with other errors. We find that tailoring the fault-tolerance framework to the physical system allows the scheme to have a higher threshold than schemes reported in literature. We present a ballistic scheme which can tolerate a 10.4\% probability of suffering photon loss in each fusion.
\end{abstract}

\maketitle

\section{Introduction}
\label{sec:intro}
In this paper we introduce fusion-based quantum computing (FBQC),  a model of universal quantum computation that is built on two primitive operations: generation of small constant-sized entangled resource states and projective entangling measurements, which we refer to as fusion. 
In particular, we explore how topological fault-tolerant quantum computation can be realized in this model.

When it comes to the design of a quantum computer, making a careful choice of the physical primitives and how they will be composed is a crucial aspect of both quantum hardware and architecture design. 
This is particularly true in fault-tolerant architectures, where a small set of operations is used repeatedly, suggesting they should be chosen with particular care. The framework of FBQC allows for the analysis of fault tolerance and logical quantum computation in a model that intrinsically accounts for many of the most important errors and physical limitations that would arise from hardware with these primitives. It is a model where the computation is directly built out of the physical primitives available. As a result, a rich dialogue can emerge between the implementation of physical operations and fault tolerant logic. 

Quantum computation based on resource state generation and fusion has been studied previously~\cite{browne_rudolph, kieling2007percolation, GSBR, auger2018fault, pant2019percolation, li2015resource, herr2018local} by mapping operations onto a model of circuit-based quantum computing or measurement-based quantum computing (MBQC)~\cite{raussendorf2006fault,Briegel_2009}. 
Mapping from one native gate set into another introduces intrinsic inefficiencies that translate to lower thresholds against physical errors, and often, complex requirements on classical processing. By using the operations of resource state generation and fusion natively, FBQC eliminates the need for any other primitives, and gives a very low depth protocol where every physical qubit is measured out almost immediately after being prepared. This significantly reduces classical processing requirements, and provides increased tolerance to both erasure and Pauli errors. 
In this paper we do not attempt to provide optimized FBQC schemes, but rather focus on pedagogical examples.
Even with these simple schemes, we see notable performance improvements over existing protocols~\cite{herr2018local, auger2018fault, li2015resource}. We consider a hardware agnostic fusion error model and demonstrate a threshold of 11.98\% against erasure in each fusion measurement, and a threshold of 1.07\% against Pauli error.

\begin{figure*}[ht]
  \includegraphics[width=0.9\textwidth]{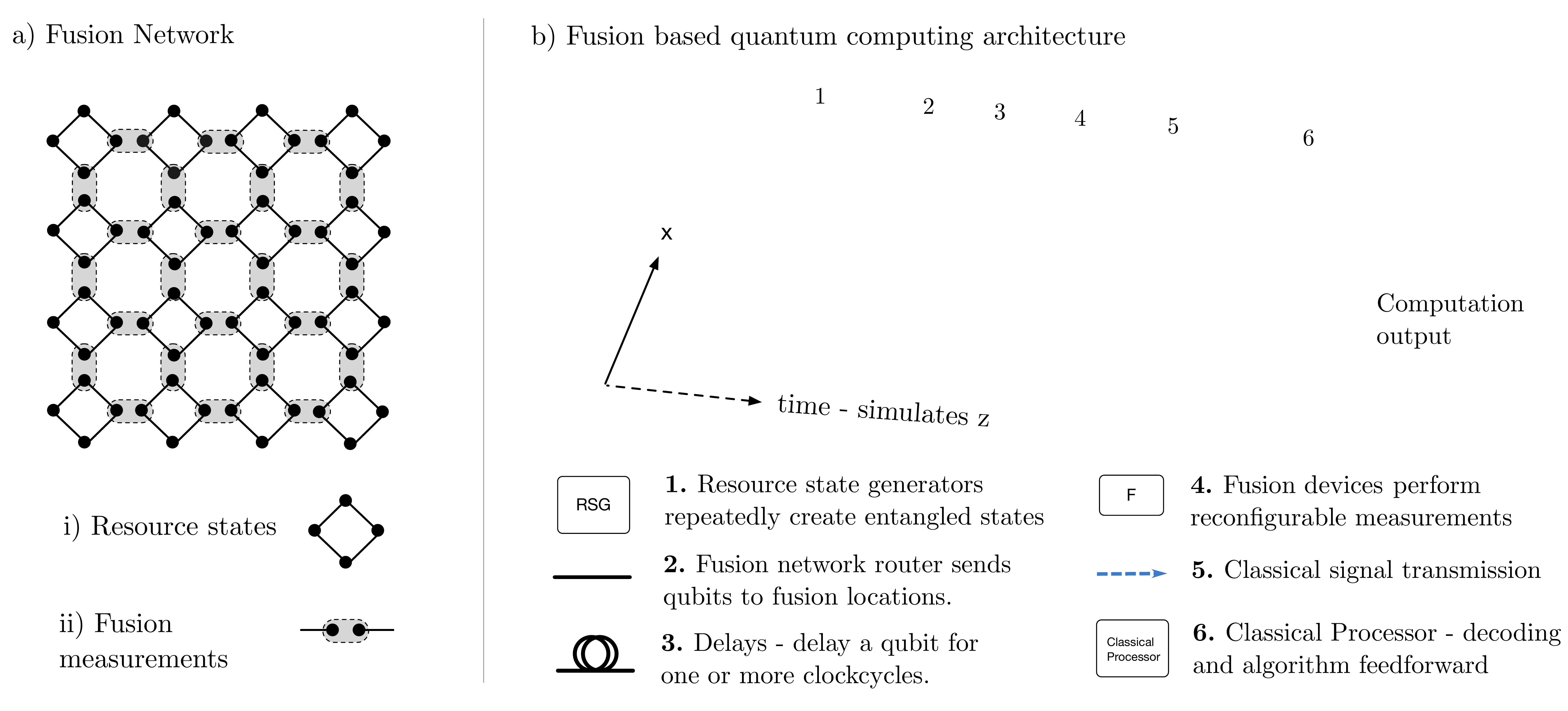}
  \caption{a) A 2D example of a fusion network, where entangled resource states and fusions are structured as a regular 2D square lattice. Resource states (i) are graph states made up of four entangled qubits in a square configuration. (ii) These qubits are measured pairwise in entangling fusion measurements as depicted by the grey shaded ovals. b) An example architecture which could create the fusion network shown in (a). Each qubit is created in a resource state generator (1) and traverses the architecture from left to right through stages labelled 2-6. Qubits are connected to fusions (4) via a fusion network router (2), which can include time delays (3). Fusion devices may be reconfigurable such that they can make projective measurements in different bases. Classical signals from fusion measurements (5) are fed to a classical processor (6), which is responsible for decoding and algorithmic logic. There can be feedforward from this computation to reconfigure fusion measurements in order to implement logic. This figure illustrates how the fusion network may include one (or more) additional dimensions compared to the hardware. Here the fusion network is 2D and the physical hardware is a 1D array of resource state generation and fusion. The physical architecture for fault tolerant computing is discussed further in section~\ref{sec:quantum-computation}.
}
  \label{fig:FN_example}
\end{figure*}

FBQC can be applied across many physical systems, and is particularly relevant to architectures where multi-qubit projective measurements are native operations. This could be linear optical fusion operations on single photons~\cite{nielsen2003quantum, browne_rudolph, kieling2007percolation, GSBR}, homodyme measurements in continuous variable photonics~\cite{gottesman2001encoding, braunstein1998teleportation,braunstein2012quantum,glancy2006error,bourassa2020blueprint}, or joint anyonic charge projection measurements in topological quantum computation \cite{Nayak_2008,chao2020optimization}. 

We focus on FBQC in linear optical quantum computing, and study the threshold behavior under a physically motivated linear optical error model accounting for photon loss as well as non-deterministic fusion operations. We show a scheme that can tolerate a 10.4\% probability of suffering photon loss in each fusion. We also demonstrate a threshold of 43.2\% against fusion failure with a ballistic scheme, compared to 14.9\% previously reported~\cite{auger2018fault}.

\subsection{Principles of FBQC}

The central principle of FBQC is to construct {\it fusion networks}, which define a configuration of {\it fusion measurements} to be made on qubits of a collection of {\it resource states}. 
The fusion network forms the fabric of the computation on which an algorithm can be implemented by modifying the basis of at least some of the fusion measurements. 

Appropriately combining fusion measurement outcomes gives the output of the computation. An example of a 2-dimensional fusion network is shown in Figure~\ref{fig:FN_example}(a). In general there is no requirement for any particular structure in the fusion network, but since our goal here is to construct topologically fault-tolerant fusion networks all the examples we look at are geometrically local.

The resource states in the fusion network are small entangled states. These states have a fixed size and structure, regardless of the size of the computation they will be used to implement. 
While these states could be any size in theory, we typically consider resource states of $O(10)$ qubits, as shown in the examples presented here. 

Fusion measurements are projective entangling measurements on multiple qubits. We consider measurements on $n$ qubits, which output $n$ classical bits giving measurement outcomes. 
For example a Bell measurement on two qubits, yielding two measurement outcomes. 
  
\subsection{Fault tolerance and Computation }

To achieve fault tolerance we must carefully choose resource states and fusion measurements and use them to construct a fault-tolerant fusion network such that the measurement outcomes combine to give parity checks of a fault tolerance scheme. In section \ref{sec:example} we give an example of a 3-dimensional topological fault-tolerant fusion network.

For a fault-tolerant quantum memory the fusion measurement configuration can be fixed, and in the example presented all the fusion measurements are in fact identical. This is similar to the stabilizer measurements that are repeatedly made in the surface code. In order to implement fault-tolerant logic, topological features such as boundaries, punctures, defects or twists must be added. All these features can be implemented simply by changing the basis of fusion measurements in a suitable location, which we discuss in Section~\ref{sec:quantum-computation}.

\subsection{Architecture}
A fusion network does not specify the ordering of operations. In the design of an architecture for FBQC we specify how the network is created from physical operations. A given fusion network has many possible architectural implementations, for example for a 3D fusion network we could choose to create all the resource states simultaneously, or instead, we could create them one 2D layer at a time. 

Resource state are repeatedly created and consumed during an FBQC protocol, and so it is helpful to consider the notion of a resource state generator, a device that repeatedly produces copies of the resource state on some clock cycle.

The physical implementation of a fusion will depend on the underlying hardware. In a linear optical system, fusion may be natively implemented by performing interferometric photon measurements between different resource states, which simply amounts to  appropriate configurations of beam splitter and  photon detectors~\cite{BrowneRudolph}. In topological quantum computation~\cite{Nayak_2008}, joint anyonic charge projection is used as a primitive for  quantum computation~\cite{Barkeshli_2016}. With GKP encoded qubits fusion can be implemented with homodyme measurements ~\cite{gottesman2001encoding}. In a matter-based system such as ion traps~\cite{review_trapped_ions} or superconducting qubits~\cite{review_superconducting_qubits}, a projective measurement can be performed using a multi-qubit entangling gate followed by single qubit measurements~\cite{molmer_sorensen}. 

The time ordering and connectivity of operations in the architecture is implemented by a {\it fusion network router} which channels qubits created at different spatial and temporal locations (i.e. from different resource state generators and time bins) to corresponding fusion locations. Thus, the fusion network router includes both spatial routing as well as temporal routing in the form of delay lines. 
Figure \ref{fig:FN_example}(b) shows a schematic example of an architecture for FBQC that creates a 2D fusion network from a 1D array of resource state generators. 
 
In certain fault-tolerant fusion networks it is sufficient for the fusion network routers to implement a fixed routing configuration. 
Fixed routing means that qubits produced from a given resource state generator will always be routed to the same location. 
This design feature is particularly appealing from a hardware perspective as it eliminates the need to be able to switch between multiple possible configurations, which may be error-prone\footnote{For example, integrated photonic components implementing fixed linear optical transformations are much higher fidelity that those that are reconfigurable.}, and reduces the burden of classical control.  

The schematic architecture in Figure~\ref{fig:FN_example}b) illustrates one of the primary advantages of FBQC from a hardware perspective: that the depth of physical operations is extremely low. Each qubit of a resource state is created, and then essentially immediately measured. This low depth is critical for minimizing accumulation of errors and tolerating leakage. 

Another simplification that can be achieved in FBQC architectures is the separation of time scales for classical control. 
As is shown in Figure \ref{fig:FN_example}, feed-forward control is required at the logical level to process and decode measurement outcomes, which then influences future logical operations. 
However, this timescale can be much longer than the clock cycle of resource state generation and fusion, and no classical computation or feedback is required on this shorter timescale. 
In other words, physical qubits do not need to wait in memory while a computation runs to decide how it should be measured.

\section{Fusion}
\label{sec:fusion}

In FBQC the initial quantum resources are small entangled resource states of a fixed size. The large scale quantum correlations necessary for universal computation are generated when we perform measurements on qubits from distinct resource states. In order for this to generate long range entanglement at least some of the measurement outcomes need to be entangling, i.e. projectors onto a subspace containing at least one entangled state. There is a long history of work~\cite{nielsen2003quantum,gottesman1999demonstrating,verstraete2004renormalization,OWQCmodel2002,leung2004quantum} exploring different types of multi-qubit measurements as a computational primitive. Here, differently to these previous works, we focus on (potentially destructive) projective measurements with only rank 1 outcomes.

In general, the measurement could be any positive operator valued measure (POVM) but, for the purposes of achieving fault tolerance, it is helpful to consider  measurements where all outcomes are projections onto stabilizer states. This makes it possible to use existing stabilizer fault tolerance methods. In the examples in this paper we focus on the case of two-qubit measurements which are Bell state projections, and we follow~\cite{browne_rudolph} in calling this \emph{Bell fusion}. Using Pauli operator notation we can describe a Bell fusion as measuring the operators $X_1X_2, Z_1Z_2$ on the two input qubits, where $X_i$ ($Z_i$) is the single qubit Pauli-$X$  ($Z$) operator on the qubit $i$. This measurement is rank 1, and the measurement operators form a stabilizer group. We will sometimes refer to this as an $\langle X_iX_j, Z_iZ_j \rangle $ fusion, where we are using the notation $\langle s_1, s_2\rangle$ to indicate a group generated by the operators $s_1, s_2$.

In the fusion networks we study here, the vast majority of fusion measurements needed to implement quantum error correction are identical Bell fusions. However, in order to implement logical gates some fraction of the measurements need to differ from the others. This can be by modifying Bell fusions to implement  different two-qubit stabilizer measurements or introducing single qubit measurements. We discuss this in more detail in Section~\ref{sec:quantum-computation}.

\subsection{Fusion in Linear Optics}
\label{sec:LO_fusion}

In linear optical quantum computing (LOQC), Bell fusion on pairs of photonic qubits is simple to perform, but does not deterministically generate entanglement. This non-determinism means that the desired two-qubit measurement outcomes are sometimes exchanged for single-qubit stabilizer measurements, an event we refer to as \emph{fusion failure}.
Architectures for LOQC must handle these non-determinstic operations.
In the FBQC schemes we describe here, these fusion failures are directly dealt with by the quantum error correction protocol.

Here we specifically consider `dual-rail' qubits composed of a single photon in two photonic modes. A photon in the first mode represents the state $\ket{0}$ and a photon in the other mode represents $\ket{1}$. This qubit encoding is attractive because it contains a fixed number of photons and so loss takes the qubit out of the computational subspace. When the qubit is measured photon detectors will count the total number of photons, and therefore a loss can be heralded, which is a powerful tool for being able to tolerate optical loss. Another advantage of the fact that all qubit states have a definite photon number in dual-rail encoding is that relative phase shifts between the optical modes of different qubits have no effect on the qubit state. This arbitrary tolerance to inter-qubit phase errors is key to the ability to store qubits in longer delay lines as shown in Figure~\ref{fig:FN_example} and discussed in more length in~\cite{interleaving}. Note that the relative phase between the two modes that make up a single dual-rail qubit must still be carefully preserved but this local control is much easier to achieve than for qubit encodings in which phase shifts between qubits can implement non-trivial operations. In the dual-rail encoding all single qubit operations can be implemented deterministically and with high accuracy with linear optical operations~\cite{politi2009integrated}. The single qubit operations we make use of here are a Hadamard operation which corresponds to a 50/50 beamsplitter and a phase gate which corresponds to a phase shift on one rail of the qubit.

\begin{figure}
    \centering
    \includegraphics[width=0.9\columnwidth]{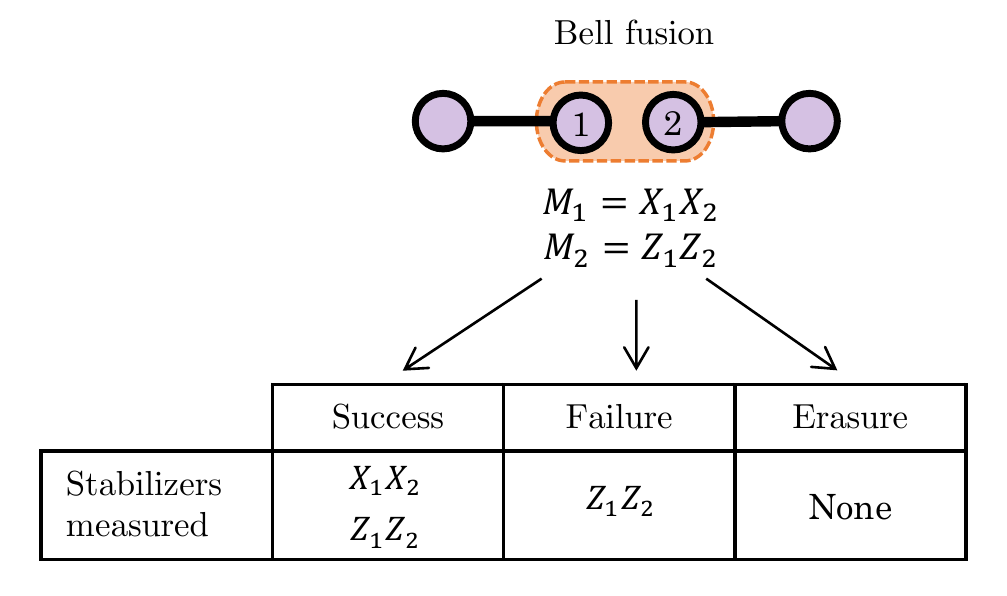}
    \caption{Outcomes of a linear optical Bell fusion. A fusion on qubits from two cluster states is shown, with intended outcomes $X_1 X_2$ and $Z_1 Z_2$. In the presence of photon loss there are three possible outcomes: \emph{fusion success} where both measurement outcomes are obtained, \emph{fusion failure} where only the outcome $Z_1 Z_2$ is obtained, and \emph{fusion erasure} where no measurement outcome is obtained. Fusion failure is intrinsic in a linear optical implementation, and happens even when all operations are ideal. Fusion erasure only occurs due to errors in the system, most commonly if one or more of the photons going into the fusion measurement are lost.}
    \label{fig:fusion_outcomes}
\end{figure}

Bell fusion on dual-rail qubits can be implemented using a linear optical circuit in which all four modes of the two qubits are measured. This is often referred to as type-II fusion~\cite{browne_rudolph}. This fusion has three possible outcomes, which are shown in Figure~\ref{fig:fusion_outcomes}. Fusion can `succeed' which happens with probability $1-p_{\rm fail}$. In this case the input qubits are measured in the Bell stabilizer basis $ X_1X_2, Z_1Z_2$ as intended. 

The fusion `fails' with probability $p_{\rm fail}$, in which case it performs separable single qubit measurements $Z_1I_2, I_1Z_2$\footnote{In \cite{browne_rudolph} the failure basis is $X_1X_2$, changing the failure basis is achieved by a simple reconfiguration of the physical device}. 
In case of such a failure, one of the two desired outcomes, $Z_1Z_2$, can still be obtained by multiplying the two single qubit measurements together. Therefore fusion failure can be treated as a Bell measurement where one of the two measurement outcomes is erased. Note that this does not generate entanglement, rather it is making use of the available classical information. In the presence of hardware imperfections (leading to photon loss and other physical level errors), there is a third possible outcome: fusion ``erasure". In this case neither of the intended stabilizer outcomes is measured.

When using dual-rail qubits with path encoding\footnote{In the dual-rail encoding, a qubit is represented by a single photon being in either of two modes. When using path encoding, the two modes correspond to two distinct waveguides, $|0\rangle$ corresponds to the photon being in the first waveguide and $|1\rangle$ to the photon in the second waveguide.}, the simplest way to implement a type-II fusion involves only two beam splitters and four detectors, and has a failure probability $p_{\rm fail} = 50\%$~\cite{browne_rudolph}. Using an additional Bell pair, fusion can be `boosted' to suppress the failure probability to $25\%$~\cite{grice2011arbitrarily,ewert20143}, and by using more ancillary photons the fusion success rate can be boosted further. It is important to note that fusion failure is a more benign error than erasure, it is not only heralded, but failed fusions produce \emph{pure} quantum states that can still be used in the computation without adding any additional noise.

We can also increase tolerance to photon loss and fusion failure by performing fusion on \emph{encoded qubits}. This method is used in the examples in section~\ref{sec:example}, where physical qubits are encoded using a [4,1,2] CSS code and encoded fusion is implemented by performing physical fusions transversally. Since this code is a concatenation of 2 repetition codes we find it convenient to refer to this as a (2,2)-Shor code where the $(N_x,N_z)$ notation indicates the number of repetitions in $X$ and $Z$ bases. In appendix~\ref{subapp:encoded_fusion_erasure_prob} we describe how erasure in an encoded fusion can be suppressed and compute the erasure probability of measurements from an encoded fusion in the presence of photon loss and fusion failure.

The effective measurement basis of linear optical fusion can be straightforwardly changed by performing single qubit rotations on one or both of the input qubits before the Bell fusion. For example, performing a Hadamard on one qubit before fusion leads to an effective fusion measurement $\langle XZ,ZX\rangle$. A switch before a fusion allows reconfigurability between different measurements, which we discuss further in appendix~\ref{app:fusion_error_model}.

\section{Resource states}
\label{sec:resource-states}

The small entangled states fueling the computation are referred to as resource states. 
Importantly, their size is independent of code distance used or the computation being performed.
This allows them to be generated by a constant number of operations. Since the resource states will be immediately measured after they are created, the total depth of operations is also constant. As a result, errors in the resource states are bounded, which is important for fault-tolerance.

Here we focus on qubit stabilizer resource states~\cite{stabilizers}, which can be described, up to local Clifford operations, by a graph $G$ using the graph state representation \cite{hein2004graph}. The graph state is defined as the quantum state $\ket{G}$ obtained by putting qubits in the $\ket{+}$ state at each vertex and performing a controlled-Z gate between qubits for which the corresponding vertices in the graph are neighbors. 
Equivalently, $n$ stabilizer generators for a graph state with vertices labeled from $1$ to $n$ are given by $X_i \prod_{j \in \mathcal{N}(i)} Z_j, i \in \{1, 2, ..., n\}$ where $\mathcal{N}(i)$ is the set of vertices neighboring vertex $i$ in $G$.

\begin{figure}
\includegraphics[width = \columnwidth]{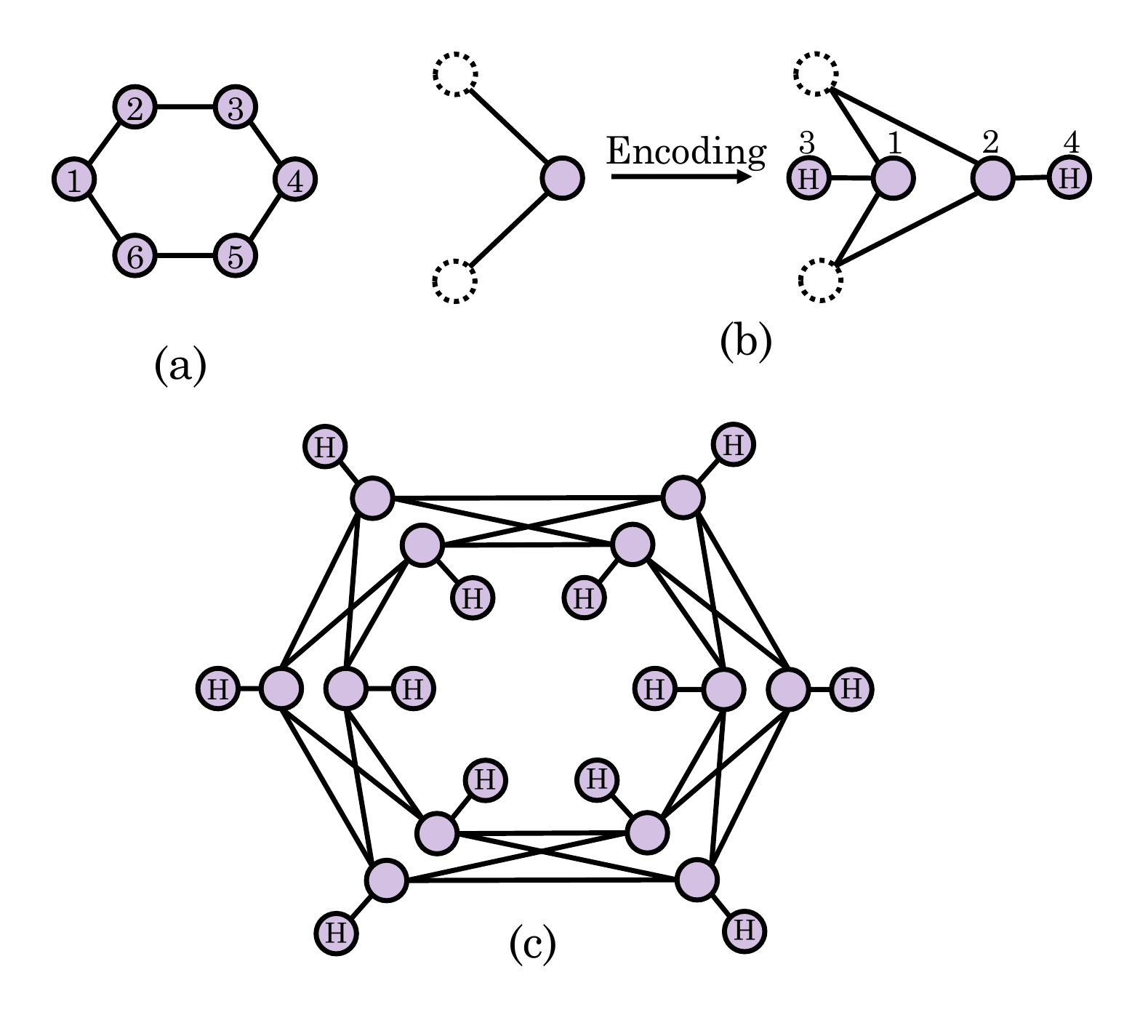}
\caption{(a) An example of a resource state represented as a graph state. 
With the qubits labelled as in the figure, the stabilizers for the resource state are $Z_6X_1Z_2$, $Z_1X_2Z_3$, $Z_2X_3Z_4$, $Z_3X_4Z_5$, $Z_4X_5Z_6$ and $Z_5X_6Z_1$. (b) A qubit in the resource state can be replaced by a (2,2)-Shor encoded resource state with the depicted transformation. The qubits 1 and 2 both have the same neighboring qubits, drawn as dotted circles, as the unencoded qubit on the left. Qubits with an $H$ inside have a Hadamard applied to them with respect to their graph state representation. (c) The resource state in (a) with every qubit encoded in a (2,2)-Shor code.}
\centering
\label{fig:resource_state_6ring}
\end{figure}

Fig.~\ref{fig:resource_state_6ring}(a) shows an example of a resource state in the form of a 6-ring graph state, where the stabilizer generators for the state are $Z_6X_1Z_2$, $Z_1X_2Z_3$, $Z_2X_3Z_4$, $Z_3X_4Z_5$, $Z_4X_5Z_6$, $Z_5X_6Z_1$. This is one of the states we use in an example of a topological fault tolerant network in Section~\ref{sec:example}. 

We also consider \emph{encoded resource states}, which can be used as a tool to reduce the impact of loss and errors in fusion measurements. In an encoded resource state the qubit operators ($X_i$, $Z_i$) are replaced by encoded qubit operators ($\bar{X}_i, \bar{Z}_i$).  We consider the example of encoding each qubit in a graph state with the (2,2)-Shor code~\cite{shor1995scheme} (as defined in Sec~\ref{sec:LO_fusion}), which has representative encoded operators: 

\begin{align}
    \bar{X}_i &= X_{i,1}X_{i,3} \\
    \bar{Z}_i &= Z_{i,1}Z_{i,2}, 
\end{align}
and each encoded qubit has stabilizers 
\begin{equation}
    S=\langle X_{i,1}X_{i,2}X_{i,3}X_{i,4}, Z_{i,1}Z_{i,3}, Z_{i,2}Z_{i,4}\rangle.
\end{equation}

This encoding is shown graphically in Figure~\ref{fig:resource_state_6ring}(b), where the qubit is replaced by 4 qubits with appropriate graph state connections to result in the redundancy of the graph state stabilizers. Replacing every qubit of a 6-ring with a (2,2)-Shor encoded qubit gives us the resource state depicted in Fig.~\ref{fig:resource_state_6ring}(c). Appendix~\ref{subapp:encoded_fusion_erasure_prob} explains the structure and effect of an encoded fusion on these states in greater detail. 

The operations used to create a resource state depend on the physical platform used, and it is worth noting that the quantum hardware used to implement fusions may differ from that used to create resource states.
In solid state qubits, for example, resource states can be generated using unitary entangling gates\cite{cirac1995quantum,molmer1999multiparticle,kane1998silicon,loss1998quantum,PhysRevA.76.042319,blais2020circuit} or dissipatively \cite{Verstraete2008, Krauter2011, Kastoryano11}.

\subsection{Resource state generation with linear optics}
When using linear optics, generation of resource states is achieved by performing a series of projective measurements, such as fusions described in section~\ref{sec:fusion}, on even smaller entangled states such as Bell states and 3-GHZ states which we sometimes refer to as {\it seed states}.
Methods for the generation of seed states are fully covered in \cite{entanglement_generation}. 
Since projective entangling measurements in linear optics succeed probabilistically, as discussed in the previous section, it is often advantageous to use switching networks between fusions to enhance the success probability of the protocol. 
Using these networks, we attempt probabilistic operations multiple times and only select cases where they have succeeded.
In this sense, multiplexing is used to effectively approximate post-selection on entangling fusion outcomes.
Since the size and number of probabilistic operations required to generate a resource state is fixed, the resource overhead from repeating probabilistic operations is constant.
There are many options for implementing such switching networks, depending on the required efficiency and available devices~\cite{switching_networks}. 
It is worth noting that to produce resource states that are qubit states, that is, states with multi-partite entanglement between well-defined qubits, the states at intermediate stages of the resource state generation do not themselves need to be qubits.

Determining the most suitable resource state is part of the design of an FBQC scheme for a realistic hardware implementation, as the noise profile of the resource state will depend on the generation protocol used. For a given target resource state there is an enormous number of possible preparation protocols, each of which will result in a different noise profile. However, the fixed size of the resource state implies that any generation protocol will require a finite number of operations, and therefore the noise accumulated in any of the state generations will be bounded. Moreover, any error correlations that emerge from independent state generation will be local to that state, which limits the spread of errors in the fusion network and is discussed in section~\ref{subsec:FBQC_errors}.

\section{Fusion networks}
\label{sec:fusion-networks}

A fusion network specifies an arrangement of resource states and a set of fusion measurements to be made on qubits of the resource states. After the measurements are made the qubits that were fused are removed from the state, and we learn measurement outcomes from each fusion. Two types of information remain: classical information from the measurement outcomes, and (potentially) some quantum correlations corresponding to unmeasured qubits. These measurement outcomes contain correlations that are the ‘outcome’ of the fusion network, giving us both a computational output or, in the case of fault-tolerant fusion networks, parity checks that can be used for error correction. 
In this section, we describe how to construct fusion networks, and how to analyze them to identify the quantum and classical correlations that exist after fusion measurements have been made. 
In particular, we focus on \emph{stabilizer} fusion networks where resource states are stabilizer states and fusion measurements are stabilizer projections.

Stabilizer fusion networks can be characterized by two Pauli subgroups: (1) A stabilizer group, $R$, describing the ideal resource states and (2) the fusion group, $F$, which is a Pauli sub-group that defines the fusion measurements, where we include $-\mathbf{1} \in F$. 
 
$F$ is not a stabilizer group since the signs of the operators are determined only after measurement (when an element of $F$ anti-commutes with an element of $R$ the outcome of the fusion measurements will be random). It is therefore convenient to include $-\mathbf{1}$ in our definition of $F$.\footnote{Defining $F$ to include $-\mathbf{1}$ is also convenient when defining the check group in Section~\ref{sec:ftfn} because a product of fusion measurements that is equal to an element of the resource state group \textit{up to signs} gives us a check.}

The result of the fusion process, in particular the relationship between the measurement outcomes and the final state of the remaining qubits, can be understood using the stabilizer formalism. As it turns out a key role is played by the \emph{surviving stabilizer group}, which is the set of elements of $R$ that commute with all elements of $F$. This is also known as the centralizer of $F$ in $R$, for which we use the following notation:
\begin{equation}
  S:=\mathcal Z_{R}(F),  
\end{equation}

\noindent Since in general $F$ and $R$ do not commute, the stabilizers after measurement will be updated. If fusion measurements are non-destructive, the new stabilizer of the final state will include the surviving stabilizer group, $S$. For an ideal fusion process, which \textit{is} destructive, the state of the \emph{outer qubits} (those remaining after measurement) is instead described by the stabilizer $S_{\rm out}$, which is the restriction of $S$ to the outer qubits up to signs. The signs are determined by the fusion measurement outcomes, and the elements of $S$ are a handy guide for determining these signs. Namely, if $s_{\rm in} \otimes s_{\rm out} \in S$, where `in'
refers to the inner (or measured) qubits, then the eigenvalue $m=\pm1$ of $s_{\rm in}$ is determined by fusion outcomes and we have:

\begin{equation}
 m s_{\rm out} \in S_{\rm out}. 
\end{equation}
In other words, the surviving stabilizer group both determines the correlations on the post-fusion state and describes the origin of those correlations.

\begin{figure}
\includegraphics[width = \columnwidth]{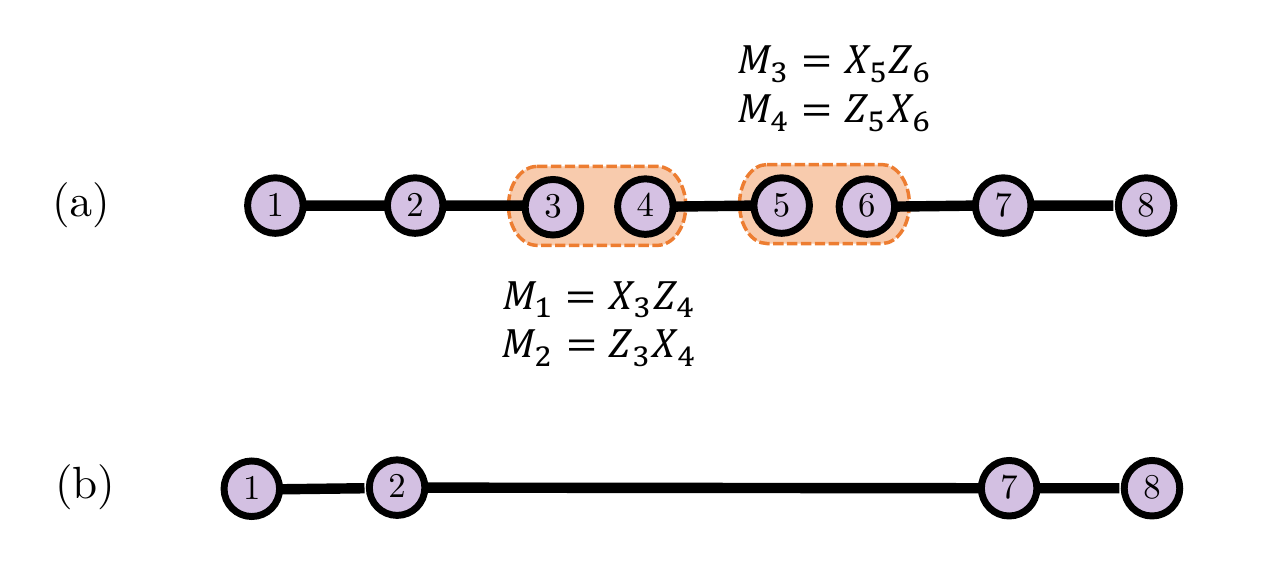}
\caption{(a) An example of a fusion network with three resource states: a two qubit graph state, and two copies of a 3-qubit linear graph state. 
There are two fusions, shown by the orange lines, both of which measure the operators $\langle XZ, ZX \rangle$. 
Specifically, the resource state composed of qubits $\{1, 2, 3\}$ is stabilized by  $\langle Z_1X_2Z_3, X_1Z_2I_3, I_1Z_2X_3 \rangle$ and similarly for $\{6, 7, 8\}$. The qubits $\{4, 5\}$ are stabilized by $\langle X_4Z_5, Z_4X_5 \rangle$. If every measurement result in the fusion network is successful and returns a $+1$ eigenvalue, the unmeasured qubits $\{1,2,7,8\}$ are stabilized by $\langle X_1Z_2, Z_1X_2Z_7, Z_2X_7Z_8, Z_7X_8\rangle$ which corresponds to the 4-line graph state shown in (b).
}
\centering
\label{fig:example_4lineRS}
\end{figure}

A simple example of a fusion network that creates a larger resource state from several smaller states is shown in Fig.~\ref{fig:example_4lineRS}(a).
The resource state group is generated by the union of the stabilizers of different resource states that can be inferred from their graph state representation:

\begin{align}
    R = \langle & (X_1Z_2, Z_1X_2Z_3, Z_2X_3),(X_4Z_5, Z_4X_5), \nonumber \\ 
                &  (X_6Z_7, Z_6X_7Z_8, Z_7X_8)\rangle,
\end{align}
where the brackets indicate the stabilizer generators associated with a single resource state. The fusion group is generated by the union of all fusion measurement operators,

\begin{equation}
F = \langle (X_3Z_4, Z_3X_4), (X_5Z_6, Z_5X_6), \mathbf{-1} \rangle, 
\end{equation}
where the brackets indicate measurement operators from the same fusion. The classical information produced by the fusion network comes from the measurement results of $F$. We use $m_i$ to denote the classical measurement outcome of the measurement operators labelled $M_i$ as shown in Figure~\ref{fig:example_4lineRS}(a).
After fusion measurements are made, the output stabilizer generators, $S_{\rm out}$, are given by: 
\begin{align}
    S_1 &= X_1 Z_2, \\
    S_2 &= m_2 m_4 Z_1 X_2 Z_7, \\
    S_3 &= m_1 m_3 Z_2 X_7 Z_8, \\
    S_4 &= Z_7 X_8,
\end{align}
where the signs of $S_2$ and $S_3$ are determined by the fusion measurement outcomes, $m_1-m_4$. The output of this fusion network therefore corresponds to the 4 qubit linear graph state shown in Fig.~\ref{fig:example_4lineRS}(b). By starting with resource states and fusions we are left with quantum correlations on the remaining qubits, with signs that are determined by measurement outcomes from the fusions in the network. This is the same principle that we will deploy to achieve fault tolerance in the next section.

It is worth noting that a fusion network as we have described it here simply represents the entanglement structure of states and measurements. It does not specify the ordering of operations, nor is it necessary that the entire fusion network exists simultaneously. When considering an architecture to implement a fusion network a specific time ordering can be introduced as a tool for architectural design, which is explored further in~\cite{interleaving}. 

In the next section, we describe how redundancy can be added to a fusion network for fault-tolerance, and in section \ref{sec:example} we present several explicit examples of fault-tolerant fusion networks.

\section{Fault-tolerant fusion networks}
\label{sec:ftfn}

Fusion networks can be constructed to be fault tolerant, such that errors in resource states, or noisy fusion measurements can be corrected for, as long as errors occur with sufficiently low probability. 
In this section we describe fault tolerance in fusion networks.

Fault-tolerant fusion networks (FTFNs) can be constructed in a way that is inspired by circuit-based quantum error correction or fault-tolerant cluster states. 
This approach can be a useful initial guide, but a direct translation often yields inefficient schemes, and better approaches can be found by working more directly in the fusion network picture, as we show in the examples in section \ref{sec:example}.

\subsection{Stabilizer formalism for FTFNs}
\label{subsec:FTFN_stabform}

The surviving stabilizer group helps us describe the output of a fusion network, notably through two of its subgroups: the output stabilizers and the check operators. 

\textbf{Output stabilizers.} The output stabilizers, $S_{\rm{out}}$, can be thought of as the logical operators of the system. In a fault-tolerant fusion network the existence of check operators allows errors in measurements to be identified and corrected, adding redundancy to the output stabilizers. 

\textbf{Check operators.} 
The key to fault tolerance is a redundancy in the fusion outcomes, which is precisely captured by the check operator group,
\begin{equation}
C:=R \cap F .
\end{equation}
The elements of this stabilizer group are the stabilizers of $R$ that are effectively measured in the fusion process. 

In the absence of errors fusion outcomes are such that the generators of the check group, $C$, have positive eigenvalues, even though the individual fusion outcomes are random. That is, the set of all possible fusion outcomes forms a binary classical code. Recall that $\mathbf{-1} \in F$ so elements of $F$ and $R$ that are equal up to signs will still result in checks.

\textbf{Errors}
When imperfections are present resource states will differ from the ideal state, resulting in Pauli errors on some qubits. The fusion measurements may also be imperfect, adding additional noise, or leading to flipped measurement outcomes. In considering these different sources it is helpful to describe all errors as an element of the Pauli group applied to the qubits in the fusion network after ideal resource state preparation, and before ideal fusion measurements. This is sufficient to capture both imperfections in resource generation and errors that occur during fusion. 

\textbf{Detectable Errors}
Detectable errors are those Pauli operators that affect one or more of the check operators, such that their measured eigenvalue is negative. If an error, $E$, anti-commutes with a fusion measurement then the value of the fusion measurement will be flipped.
The value of all check operators is referred to as the \emph{syndrome}. 
The classical fusion outcomes form a classical linear binary code, which allows them to be corrected. Given the syndrome, it is the job of a decoder to predict the sign of the output stabilizer based on the most likely physical error class consistent with the detection pattern. If the decoder predicts the error class correctly, all logical outcomes will be recovered correctly. 
Note that the error correction being performed here is entirely on the classical fusion measurement outcomes, and the output of decoding is used to update the Pauli frame that allows us to interpret the remaining state. There is no notion of a physical correction that could be applied, since the qubits that suffered errors have been destructively measured.

\begin{figure*}[ht]
\includegraphics[width = 2.1\columnwidth]{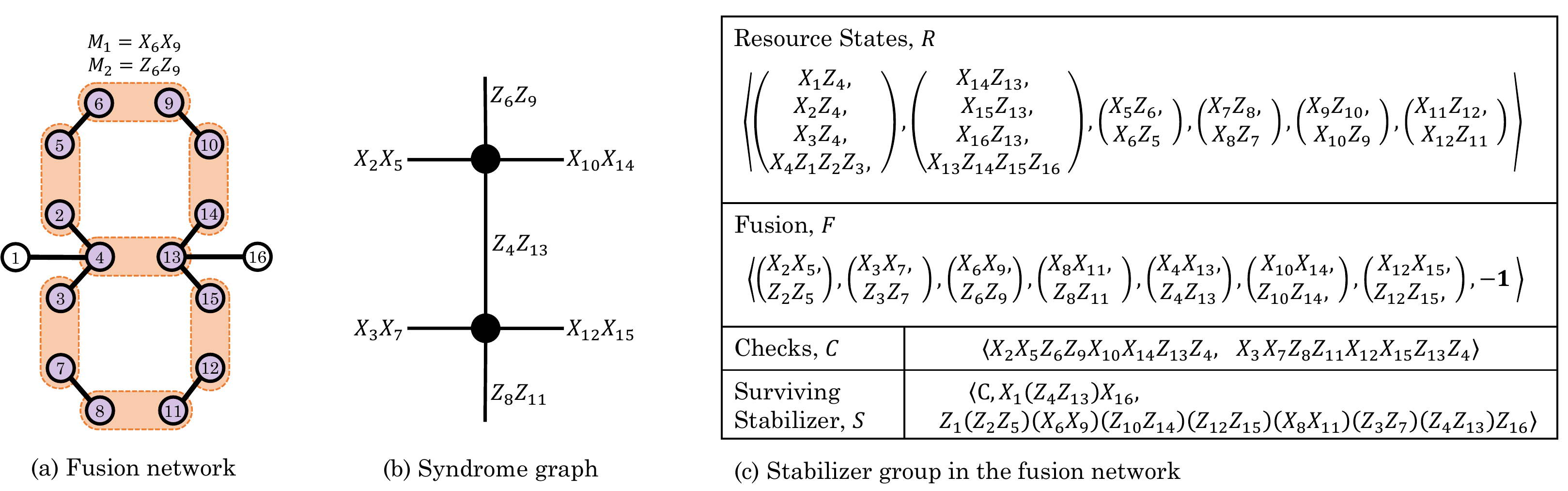}
\caption{ Example fault-tolerant fusion network. (a) Arrangement of resource states, represented as graph states. Qubits that will be measured during fusion are shaded, and the output qubits that remain after measurement are unshaded. 
Fusion measurements are shown by orange ovals. All fusions are of the type $\langle X_i X_j, Z_i Z_j \rangle$.  (b) 
The syndrome graph corresponding to (a) with measurements corresponding to every edge labelled. Multiplying the measurement operators adjacent to the vertices gives us the generators of the check group, $C$. (c) Explicit definition of the generators of each stabilizer group, where the brackets in the expressions are a guide to indicate the separable sub-groups of $F$ and $R$.  $R$ is the union of the stabilizers of different resource states that can be inferred from their graph state representation. The fusion group is generated by all the fusion  measurements and $-\mathbf{1}$. Generators from different resources states in $R$ and different fusions in $F$ are sorted by column. The surviving stabilizer group, $S$, includes check operators, and stabilizers that include the output qubits (1,16). 
}
\centering
\label{fig:example_FTFN}
\end{figure*}

\textbf{Undetectable errors} 
The group of undetectable errors is defined by the centralizer of the check group $C$ on the whole Pauli group:

\begin{equation}
U:=\mathcal Z(C).
\end{equation}
As the name suggests, this is the subgroup of the Pauli operators which leaves no trace on the check operator results.
However, not all undetectable errors are problematic for computation, since some do not affect the final correlations of interest.
For example, $U$ contains elements of $R$ (or $F$) which will have no detrimental effect as they leave the resource states (or fusions) invariant.
More generally, the group of trivial undetectable errors is
\begin{equation}
T:=\mathcal Z(S),
\end{equation}
where $S$ is the surviving stabilizer group defined in section~\ref{sec:fusion-networks}. $T$ includes $\langle R,F \rangle$ by definition. Undetectable errors can thus be classified by the elements of the quotient $U/T$. Trivial errors, that are elements of $T$, have no effect on the check operators, or the output stabilizers. Errors may also be non-trivial but undetectable. These errors do not affect the check operators, but do affect the output stabilizers. 
For fault-tolerance, we are interested in fusion networks where the weight of non-trivial undetectable errors, also called the distance of the code, increases with the size of the network.
Some examples of such networks are discussed in section~\ref{sec:example}.

\textbf{Error representation.} Errors may occur in the generation of resource states, or in the fusion measurements themselves. It is convenient to choose a representation that groups together errors which have an equivalent action on the code state and check operators.  It is never necessary to distinguish different errors which are equivalent up to an element of $F$, since $F \subseteq T$. We therefore choose to express decoding problems in terms of elements of $P/F$, where $P$ is the Pauli group (i.e. the full Pauli group quotiented by the fusion group).
While distinct elements of $P/F$ may correspond to equivalent errors (according to the fully reduced equivalence classes of $P/T$), the partial reduction $P/F$, has the advantage of preserving a large amount of the locality structure in the error model.
In particular, single qubit Pauli errors on resource states are interpreted as measurement errors on a corresponding generator(s) of $F$.
When $F$ is composed of Bell fusion measurements, the quotient $P/F$ identifies pairs of single qubit Paulis in $P$ whenever they jointly produce an element of $F$.
We can thus choose to express decoding problems in terms of $P/F$, which directly corresponds to specifying which fusion outcomes were flipped.
For instance, in the example of Fig.~\ref{fig:example_FTFN}, the single qubit errors $X_4$ and $X_{13}$ are equivalent errors, as they multiply to $X_4X_{13}$, which is an element of $F$.
Up to this equivalence, errors can be characterized by which generators of $F$ (fusion outcomes) they flip, in this case $Z_4Z_{13}$.

\subsection{Example fault-tolerant fusion network}

Figure~\ref{fig:example_FTFN} shows an example of a fusion network, where the output is a Bell pair on qubits 1 and 16. The stabilizer groups $R,F,S$ and $C$ for this example are defined in the table in Fig.~\ref{fig:example_FTFN}(c). This network contains two check operators, which allow certain errors in the fusion network to be corrected.

After all fusions have been performed the quantum correlations of the unmeasured qubits depend on the classical measurement information, and this dependency is captured by the surviving stabilizer group as we saw in section~\ref{sec:fusion-networks}. We use the notation $m^{XX}_{i,j}$ to represent the classical bit from the measurement of the operator $X_iX_j$ and $m^{ZZ}_{i,j}$ for the measurement of $Z_iZ_j$.

The outcomes (eigenvalues) of the check operator generators are:

\begin{align}
    C = \langle & m^{XX}_{2,5} m^{ZZ}_{6,9} m^{XX}_{10,14} m^{ZZ}_{4,13},\nonumber \\
                & m^{XX}_{3,7} m^{ZZ}_{8,11} m^{XX}_{12,15} m^{ZZ}_{4,13}\rangle \label{eq:example_check}.
\end{align}
Since check operators should have positive eigenvalue, they act as parity checks on the bits describing the fusion outcomes. The output stabilizer group is: 

\begin{align}
    S_{\rm out} = \langle & m^{ZZ}_{4,13} X_1X_{16}, \\
                    & m^{XX}_{4,13} m^{ZZ}_{2,5} m^{XX}_{6,9} m^{ZZ}_{10,14} m^{XX}_{12,15} m^{XX}_{8,11}m^{XX}_{3,7} Z_1Z_{16} \rangle \nonumber
\end{align}
such that when all fusions succeed the fusion network produces a Bell pair $\langle \pm X_1X_{16}, \pm Z_1Z_{16} \rangle$. 

In this small example some (but not all) fusion measurement errors can be corrected. For example, consider a measurement error in $Z_4Z_{13}$. The corresponding outcome, $m^{ZZ}_{4,13}$ appears in the output stabilizer, determining the sign of the $XX$ stabilizer of the output Bell state, so an error will result in an incorrect sign. However, the measurement also appears in both checks in $C$, and so it can be corrected. 

In this example we can also identify the other types of error configuration. Trivial errors  correspond to elements of $T$, for example an error $E_1 = Z_2Z_{10}$ will flip measurement outcomes $X_2X_5$ and $X_{10}X_{14}$ but neither check is affected. Since $E_1 \in R \cup F$ this has a trivial effect on the output stabilizer. 

Consider another error: $E_2 = Z_4$. This is a non-trivial undetectable error, which causes a single flipped measurement in the fusion outcome $X_4X_{13}$. This error commutes with the check group $C$, which we can see by noting that $m^{XX}_{4,13}$ is not present in either of the check operators, and it will therefore go undetected. $E_2$ also anti-commutes with one of the additional generators in $S$, since we can see its presence in the output stabilizer. For this reason, this error will lead to an incorrect prediction for the sign on $\pm Z_1Z_{16}$ on the output stabilizers.

\subsection{Topological fault-tolerant fusion networks}

We can introduce a notion of low-density parity check (LDPC) fusion networks that is analogous to other similar constructions in error correction~\cite{tillich2013quantum, gottesman2013fault}. Namely, a family of stabilizer FTFNs is LDPC if:
\begin{enumerate}
    \item it has check operator generators such that each generator involves a bounded number of fusions, and each fusion is involved on a bounded number of generators, 
    \item for any integer $d$, there exist fusion networks in the family such that non-trivial undetectable errors have support on at least $d$ qubits.
\end{enumerate}
One can then apply the usual combinatorial arguments to show the existence of error thresholds~\cite{aharonov2008fault,dennis2002topological, bombin2015single}.  In the FBQC setting the fusion measurement outcomes will form a classical LDPC code. 

\begin{figure}[t]
    \centering
    \includegraphics[width = 0.75\columnwidth]{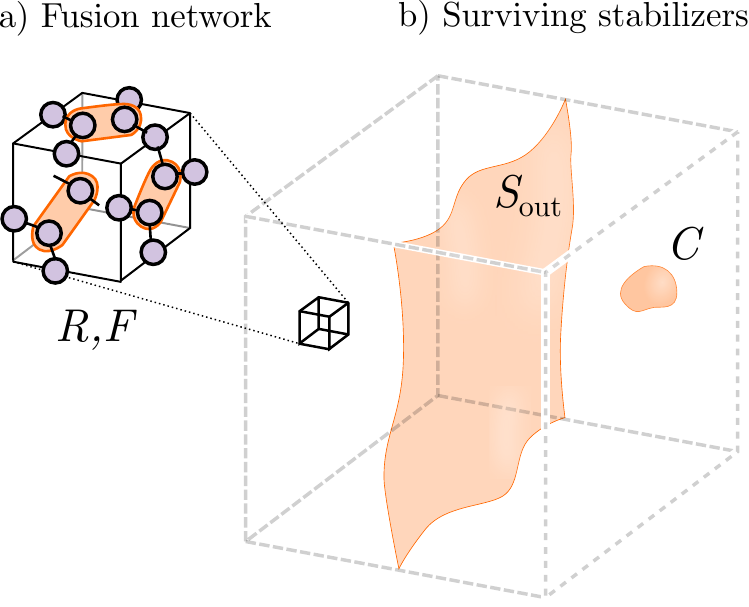}
    \caption{Illustration of the groups in a 3D topological fault-tolerant fusion network. Resource state group, R; Fusion group, F; Check group, C; Output statbilizer, $S_{\rm{out}}$ (a) Microscopically the fusion network is composed of resource states and fusions, which may form repeating unit cells. (b) After the measurements of the network are made we are left with a 3D block of classical information, with remaining qubits on the surface of the block. The check operators form closed surfaces in the bulk. The remaining generators of the surviving stabilizer group represent extensive membranes that span the network, these behave as logical operators. }
    \label{fig:topological_FTFN}
\end{figure}

We are specifically interested here in topological LDPC fusion networks, where the operations are geometrically local, and the resulting parity checks and logical operators have the structure of fault-tolerant topological error correction. The fusion network captures the behavior of a code evolving over time, so that for 2D topological codes one obtains 3D topological fusion networks. The elements of the surviving stabilizer group $S$ take the form of membranes (word-lines of string operators)  and undetectable errors in $U$ take the form of closed strings (word-lines of topological charges). An illustration of such a 3D fusion network is shown in Figure~\ref{fig:topological_FTFN}. The check operators are closed membranes, while the logical operators are open membranes that span the system and are supported on the output qubits. The introduction of appropriate boundary conditions to the fusion network allows these membranes to terminate, and we discuss this further in Section~\ref{sec:quantum-computation}. 
While there exists the noted equivalence between these 3D topological fusion networks and 2D topological codes, the 3D topological fusion networks are not constrained to represent foliated codes~\cite{foliated_codes}, in which there is a notion of a fixed code structure that persists throughout the computation. FBQC can support the more general framework of fault tolerance beyond foliation described in~\cite{nickerson2018measurement,newman2020generating}. Examples of topological fusion networks are given in section \ref{sec:quantum-computation}.

\subsection{Syndrome graphs}
\label{subsec:syndrome_graphs}

The codes associated to fault-tolerant fusion networks are often well described by a \emph{syndrome graph} representation. The syndrome graph picture makes it easier to interpret the underlying structure of the code and enables the application of existing decoders such as minimum-weight matching and union-find decoders \cite{Dennis_2002, union_find}. 
However, it should be noted that not all schemes can be naturally represented by a syndrome graph structure. That this is possible in topological (surface-code based) FBQC, is a manifestation of the fact that error chains leave non-trivial syndromes only at their endpoints.

In the syndrome graph we represent the code through a multi-graph, with vertices corresponding to check operator generators in $C$, and each edge corresponding to a generator of the fusion group $F$. An edge is connected to a vertex if the corresponding generator of $F$ is a factor of the corresponding check operator in $C$.

The parity for a check operator is evaluated by taking the joint parity of all the measurement outcomes composing the check.
Given a set of fusion measurement outcomes each parity check has an associated parity value of either +1 or -1. 
The configuration of all of these parity outcomes is called the \emph{syndrome}. 
If a fusion outcome is flipped, the vertices (checks) connected by its edge in the graph will have their parity values flipped. 
If a fusion outcome is erased, the two checks connected by the edge in the graph can be combined into a single check operator.
The syndrome graph may have `dangling edges', where an edge connects to only one check vertex. In some cases there may also be multi-edges between two check nodes when multiple fusion measurement outcomes contribute to the same pair of syndromes.

Figure~\ref{fig:example_FTFN}(b) shows how the fusion network example can be represented as a syndrome graph. 
In this simple example there are two check operators, each check operator has four incident edges. 
There is one edge that is shared by both check operators, connecting the two vertices. 
The other edges are `dangling edges' connected only to a single check node.  
In this small example not all fusion outcomes are included in a check operator, but in the examples of topological fault-tolerant fusion networks we present in the next section, all fusion outcomes will be part of at least one check operator. 
Specifically, we will consider topological fusion networks based on the surface code. 
Like any surface code construction, these networks can be represented by a \emph{primal syndrome graph} and a \emph{dual syndrome graph} which are locally disconnected. 
In the bulk, every Bell fusion contributes two measurement outcomes, which respectively correspond to one edge of the primal and one of the dual syndrome graph.

\section{Example Fault-Tolerant Fusion Networks}
\label{sec:example}

\begin{figure*}
\includegraphics[width = 1.8\columnwidth]{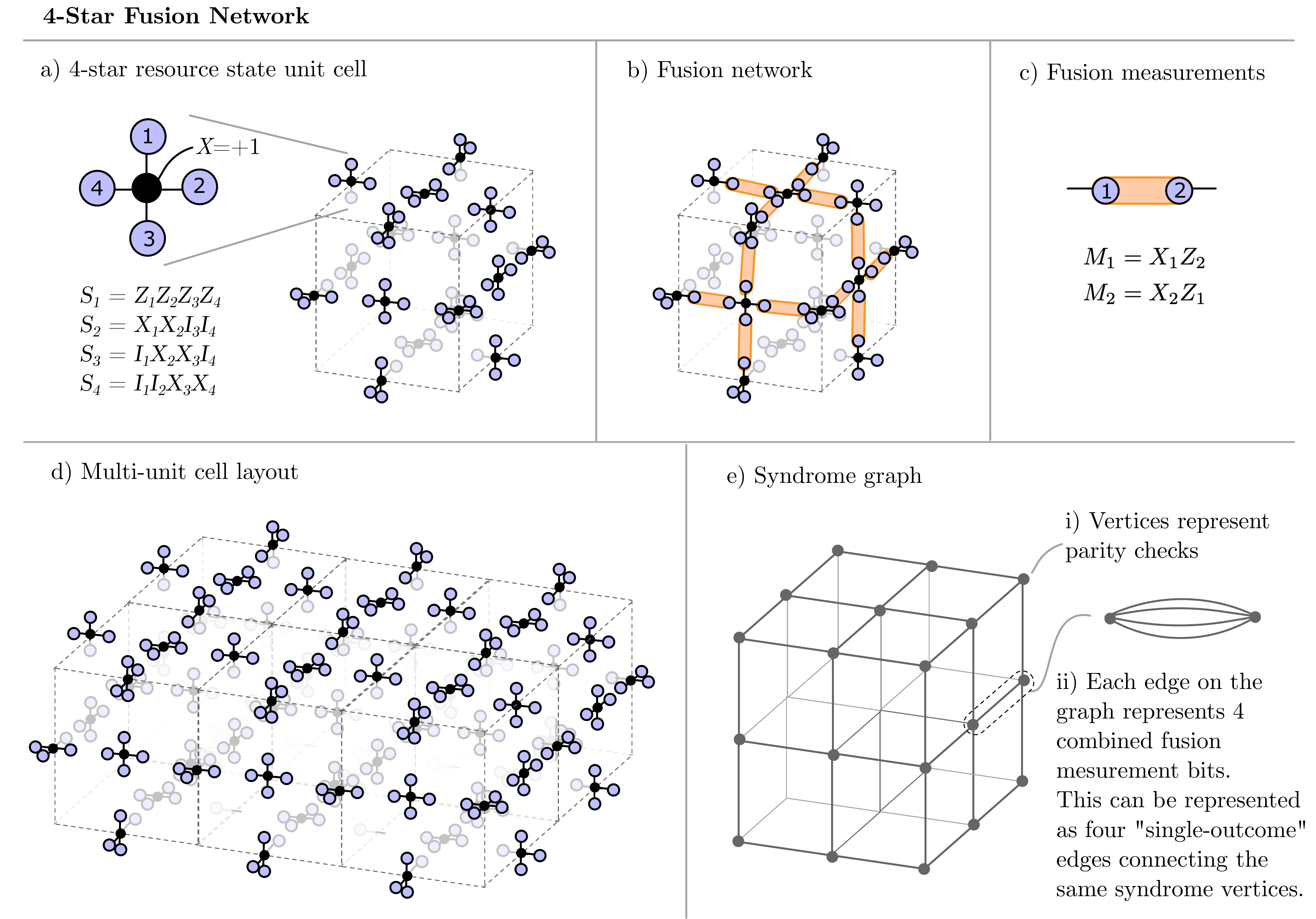}
\caption{The `4-star' fusion network: (a) Layout of resource states in a cubic unit cell of the lattice (grey dashed line). The resource state is a 4-GHZ state with stabilizers $Z_1Z_2Z_3Z_4$, $X_1X_2$, $X_2X_3$, $X_3X_4$, which is the same as a graph state in the form of a 4-star where the central qubit (shown in black) has been measured in the $X$ basis and returned a `+1' outcome. The resource state is drawn in this way only for ease of depiction; the architecture only requires 4-GHZ resource states which, in the case of a linear optical implementation, can be created directly from single photons using circuits described in \cite{entanglement_generation} . 
(b) Fusions within a unit cell shown in orange
(c) All fusion measurements in the fusion network attempt to measure $M_1 = X_1Z_2$ and $M_2 = Z_1X_2$.
(d) Layout of resource states across multiple unit cells.
(e) Syndrome graph resulting from the fusion layout. Both the primal and dual syndrome graphs have the same structure. 
Primal and dual checks can respectively be associated to unit cubes and vertices of the dashed lattice of (d).
Individual checks can be obtained by taking the product of 24 fusions between adjacent edge-face pairs (whether $XZ$ or $ZX$ is used for respectively for primal or dual).
}
\centering
\label{fig:star_layout}
\end{figure*}

In this section we describe two explicit examples of fault-tolerant fusion networks that implement surface code error correction. The first model is closely related to existing schemes in the literature for constructing cluster states for fault tolerant MBQC \cite{herr2018local, auger2018fault} while the second example shows a novel scheme which follows design principles to minimize the vertex degree of the syndrome graph.
These examples provide simple illustrations of how fault-tolerance can be achieved in the FBQC framework. They are chosen as helpful pedagogical examples and are not optimal FBQC architectures. However, even with these examples we demonstrate a significant performance improvement. 

\subsection{4-star fusion network}

The `4-star' fusion network is shown in Figure~\ref{fig:star_layout}.
The resource state is the four qubit Greenberger-Horne-Zeilinger (GHZ) state\footnote{For convenience in defining the fusion network our resource state here has the $X$- and $Z$- operators swapped compared to the usual GHZ state definition} with stabilizer generators $\langle Z_1Z_2Z_3Z_4, X_1X_2, X_2X_3, X_3X_4 \rangle$. 
For graphical clarity, we represent this resource state as a 5-qubit star graph state with the central qubit blacked out (Fig.~\ref{fig:star_layout}(a)) because this is the state obtained on measuring the central qubit of a 5-star graph state in the $X$ basis with a `+1' outcome. There is no need to prepare a 5 qubit physical resource state, the 4-GHZ state can be created directly\footnote{Using linear optics a 4-GHZ state can be prepared from single photons, using a variety of methods, such as those described in \cite{entanglement_generation}.}.
The four shaded circles, correspond to qubits in the resource state and are input to fusions in the network.
The fusion network can be built up from a cubic unit cell as shown in Figure~\ref{fig:star_layout}(a) and (d), where a resource state is placed on every face and edge of the unit cell. 
Resource states are aligned parallel to faces or perpendicular to edges. 
A fusion measurement is made on pairs of qubits from resource states centered at unit cell faces and qubits from resource states centered at neighboring edges as shown in Fig.~\ref{fig:star_layout}(b)\footnote{If the center qubit in the resource state had not been measured, the fusions would result in the cluster state used in the MBQC implementation of the surface code~\cite{herr2018local, auger2018fault}.}.
Each fusion attempts to measure the stabilizer operators $X_1Z_2$ and $Z_1X_2$ as shown in Fig.~\ref{fig:star_layout}(c).  We include a formal definition of the layout in Appendix~\ref{app:formal_fusion_network_definitions}.

For each unit cell there is a primal check operator that is associated with the cell, which is made up of 24 fusion measurement outcomes. There are 4 fusions per face of this cell, giving a total of 24 by combining all 6 faces of the cube. For each of these fusions, one of the two measurement outcomes contributes to the cell parity check. Specifically the parity check operator is $C_c = \prod_{f\in c} \prod_{e\in f} Z_f X_e$, where $c$ is the cell, $f$ is a face, and $e$ is an edge. The other measurement outcomes contribute to one of the dual parity checks, which are associated with the corner vertices of the unit cell. The fusion network is symmetric under translation by half the lattice constant in all three dimensions. This means that the primal syndrome graph and dual syndrome graph are identical.

The logical operators are defined by 2D membranes made up of connected faces of the lattice. To evaluate the membrane we combine the fusion measurement outcomes from all four fusions on each face that makes up the membrane. Specifically the operator for a membrane, $\mathcal{M}$, is defined as $\prod_{f\in \mathcal{M}} \prod_{e \in f} Z_f X_e $.

The fusion network results in the syndrome graph shown in Fig.~\ref{fig:star_layout}(e): a cubic lattice, where every edge is a multi-edge corresponding to four measurement outcomes. In total there are 24 fusion measurements that combine to evaluate each check operator. Since the syndrome graph is used across different implementations of the surface code, it is a useful tool to understand the correspondence between FBQC and a circuit based surface code model. In a circuit model, space-like edges in the 3D syndrome graph correspond to physical qubit errors, while time-like edges correspond to measurement errors. 
In FBQC both time-like and space-like edges correspond to fusion measurement outcomes - there is no distinction between physical and measurement errors in this model. Another point of comparison is the interpretation of the primal and dual syndrome graphs. In a circuit model the primal syndrome graph captures Pauli-X errors, and measurement errors on Z-type parity checks, while the dual syndrome graph captures Pauli-Z errors and measurement errors on X-type checks. 
In this FBQC example, each 2-qubit fusion contributes one measurement outcome to the primal graph, and the other to the dual graph. 
One way of viewing this is that the two fusion measurement outcomes behave like the Pauli-X and Pauli-Z parts of the error channel on a physical qubit in the circuit model. 

For a fusion network like this, that generates the bulk of a fault tolerance structure, it is always possible to swap the resource states and measurements to find another valid fault tolerant fusion network, i.e. defining $R' = M$ and $M' = R$. In this case that would mean that the resource states become bell pairs, and the fusion measurements become 4-qubit projective measurements on a GHZ state. Depending on the hardware available, one or the other of these fusion networks may be preferable.

\subsection{6-ring fusion network}\label{sec:example_6ring}

\begin{figure*}
\includegraphics[width =1.8\columnwidth]{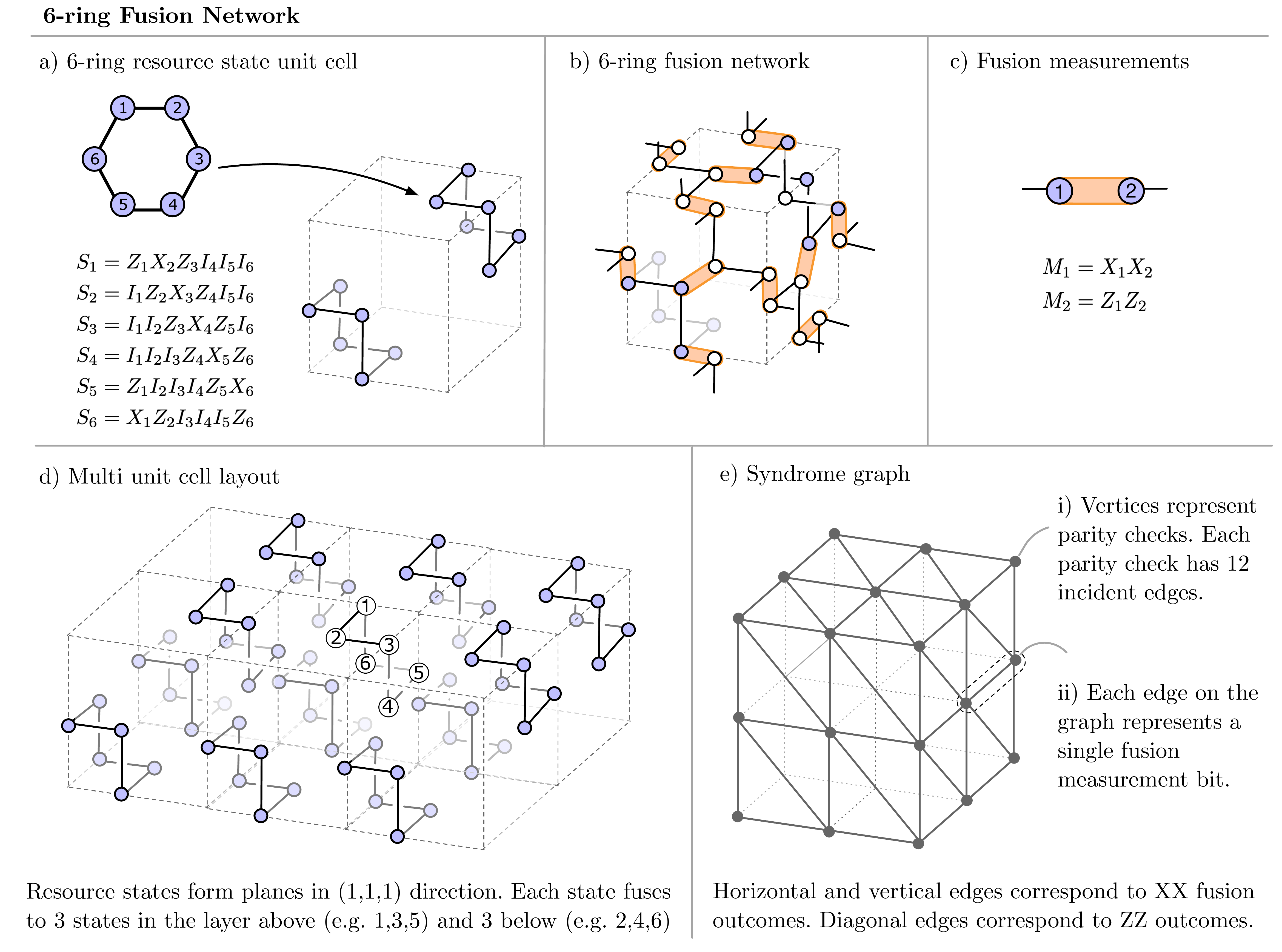}
\caption{The `6-ring' fusion network. 
(a) Each resource state is a graph state in the form of a ring of six qubits. Two resource states are placed at opposite corners of each unit cell. (b) Two-qubit fusions connect every pair of qubits that share a face or an edge. Resource states that belong to the unit cell are shown as shaded circles, while qubits from resource states in neighboring cells are shown as white circles. A formal definition of the fusion network can be found in Appendix~\ref{app:formal_fusion_network_definitions}. 
(c) All fusion measurements in the fusion network are two-qubit projective measurements on the bases $M_1 = X_1X_2$ and $M_2 = Z_1Z_2$.
(d) Shows the layout of resource states across multiple unit cells. When unit cells are tiled the resource states can be grouped into layers along 2D planes perpendicular to the (1,1,1) direction. 
Three qubits in each state fuse with the layer above, and three with the layer below. 
(e) The syndrome graph resulting from the fusion layout is a cubic graph with diagonal edges as shown. 
Primal and dual syndrome graphs have an identical structure. 
In both, the vertical edges correspond to $XX$ type fusion outcomes and diagonal edges correspond to $ZZ$ outcomes.
}
\centering
\label{fig:kagome6_layout}
\end{figure*}

Our second example, the 6-ring fusion network, improves on the 4-star network. It requires fewer resource states and fewer fusion measurements to implement a code of the same distance, and as we will see in the next section, it offers a significantly improved threshold.  

In this fusion network, the resource states are graph states in the form of six qubit rings, with stabilizers $\langle Z_1X_2Z_3, Z_2X_3Z_4, Z_3X_4Z_5, Z_4X_5Z_6, Z_5X_6Z_1, Z_6X_1Z_2\rangle$. The fusion network has a cubic unit cell with two resource states per unit cell, as depicted in Fig.~\ref{fig:kagome6_layout}(a). Fusion measurements connect the pair of qubits at each face and each edge, as shown by the orange lines in Fig.~\ref{fig:kagome6_layout}(b).
Each fusion attempts to measure the stabilizer operators $X_1X_2$ and $Z_1Z_2$ on the input qubits. Figure~\ref{fig:kagome6_layout}(d) shows multiple unit cells, where it can be seen that the resource states form layers in the planes perpendicular to the (1,1,1) direction. Appendix~\ref{app:formal_fusion_network_definitions} gives a formal definition of the 6-ring fusion network. 

There are 12 fusion measurements that combine to evaluate each check, which is half as many measurements per check compared to the 4-star network. For each unit cell there is a primal check operator that is associated with the cell. For each cell, the check operator comprises measurements for one fusion per face, and additionally fusions for 6 edges of the cell. The edges that contribute to the cell’s check are the 6 edges that do not have a qubit from a resource state that belongs to the cell, as represented in Fig.~\ref{fig:kagome6_layout}(a). At these edges a fusion measurement is needed to link together the stabilizers of neighboring faces, we call this set of edges $E_{\rm{link}, c}$. For each of these fusions, one of the two measurement outcomes contributes to the cell parity check. Specifically the parity check operator is $C_c = \prod_{f\in c} (XX)_f \prod_{e \in E_{\rm{link},c}} (ZZ)_e$, where $c$ is the cell, $f$ is a face, and $e$ is an edge. The other measurement outcomes contribute to one of the dual parity checks, which are associated with the corner vertices of the unit cell.

The logical operators are defined by 2D membranes made up of connected faces of the lattice. To evaluate the membrane we combine the fusion measurement outcomes from the fusion on each face, and the fusion at each connection between two faces. For example if we consider a flat 2D sheet membrane, $\mathcal{M}$, then we can define the membrane operator as $\prod_{f\in \mathcal{M}} (XX)_f \prod_{e \in \mathcal{M}} (ZZ)_e $. Other representations of the logical membrane can be obtained by multiplying by check operators.  In some configurations two connected faces of the membrane may be connected through a resource state, in which can the fusion at this edge is not included in the membrane operator.

The syndrome graph for this fusion network is depicted in  Fig.~\ref{fig:kagome6_layout}(e), and is a cubic lattice with added diagonal edges. The 6-ring network has the same symmetry under translation in all three dimensions by half the lattice constant. So, as in the 4-star network, the primal and dual syndrome graphs are identical. 

The diagonal edges that show up in the syndrome graph here are a familiar feature in circuit based surface codes, where they would be interpreted as so-called `hook' errors. These can occur when a single error event spreads to neighboring qubits during the stabilizer measurement circuit. Although the origin of this type of correlated error is very different in the fusion network setting, in both cases these edges arise due to the process of creating large scale entanglement from physically low weight operations.

\subsection{Performance comparison}

We study the performance of these two fusion networks by simulating their behavior under a model of both Pauli errors and erasure. We consider two error models:

\begin{itemize}
    \item A hardware-agnostic fusion error model where every fusion measurement is erased with some probability, $p_{\rm{erasure}}$, and flipped with some probability, $p_{\rm{error}}$. 
    \item A linear optical error model where every fusion has a failure probability, $p_{\rm{fail}}$, and every photon in a resource state has a probability of being lost, $p_{\rm{loss}}$.
\end{itemize}

\subsubsection{Hardware-agnostic fusion error model}

\begin{figure}
\includegraphics[trim = 0 0 40 30 , clip, width=\columnwidth]{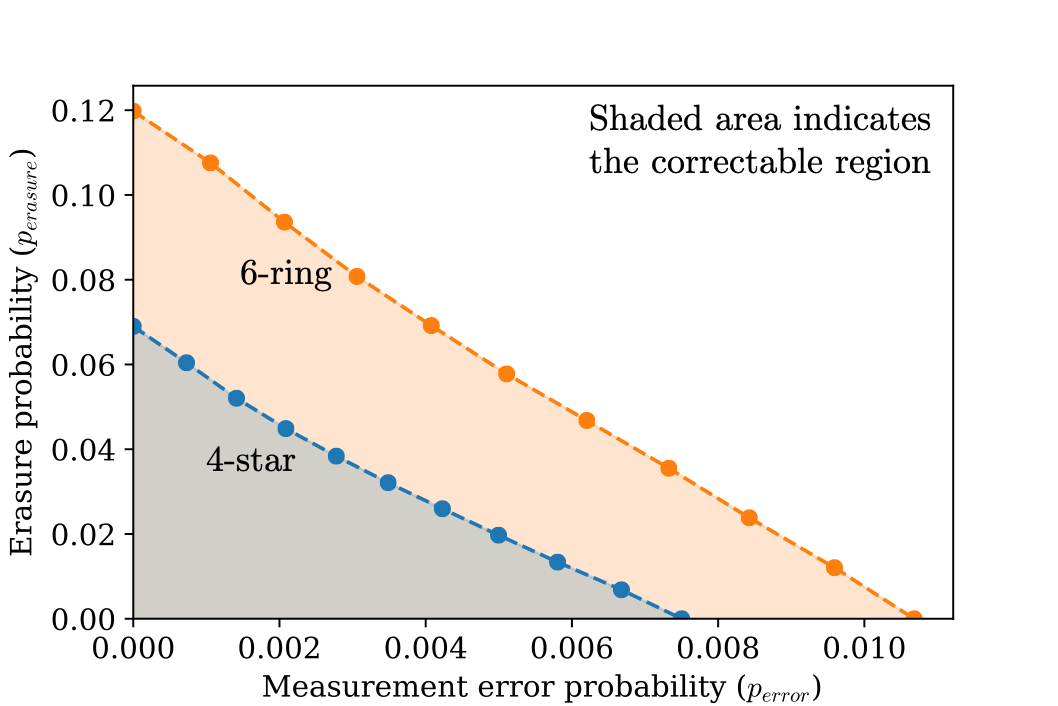}
\caption{
Performance of the 4-star (blue line) and six-ring (orange line) fusion networks. The correctable region is shown for the two fusion networks under the two error parameters of the hardware-agnostic fusion error model: fusion erasure probability  $p_{\rm{erasure}}$ and measurement error probability $p_{\rm{error}}$. Each marker shows the position of the threshold in the 2 parameter space, and is evaluated by a series of Montecarlo error sampling and decoding trials at different error parameters.  Simulation details are provided in Appendix~\ref{sec:simulation_methods}.
}
\label{fig:example_phenomenological_phasediagram}
\end{figure}

Every fusion in a fusion network produces two measurement outcomes which we call \textit{fusion measurements}. In the hardware-agnostic fusion error model every measurement in the fusion network is erased with probability $p_{\rm{erasure}}$ and flipped with probability $p_{\rm{error}}$.
This can capture single qubit Pauli errors and erasures originating from the resource state generation as well as those resulting from noisy fusion measurements, as discussed in Section~\ref{subsec:FTFN_stabform}, since they can all be characterized by the fusion outcomes that they flip. Leakage and fusion failure lead to the erasure of fusion outcomes. This model is totally general in the sense that it captures some effect from all possible sources of error, but simplified in that each fusion is treated as identical and bit-flip and erasure errors are assumed to be independent.

Compared to previous studies of fault-tolerant measurement based quantum computing (MBQC), which look at the erasure and error thresholds of single qubit measurements on large cluster states which already have long range entanglement\cite{raussendorf2006fault,barrett2010fault,nickerson2018measurement}, this model captures errors in the joint measurements used to create long range entanglement starting from small resource states.
In this way, this fusion error model is closer to a circuit level error model where individual resource states and fusion measurements play the role of elementary gates. We discuss the scope and limitations of this error model in Section~\ref{subsec:FBQC_errors}.

To evaluate the threshold of each fusion network we perform Monte Carlo simulations of each fusion network with $L\times L \times L$ unit cells and with periodic boundary conditions in all three dimensions. We draw error samples based on the models above, and use a minimum weight perfect matching decoder~\cite{edmonds_1965,kolmogorov2009blossom} to perform decoding and count the instances of logical errors. Primal and dual syndrome graphs are decoded separately, but we identify a logical error as a non-trivial error chain in either the primal or dual, and in any of the three dimensions of the fusion network. We repeat these simulations over a range of values of different error parameters, and the system size, $L$, in order to build up a threshold surface in the two error parameters.  More details on our simulation methods are given in Appendix~\ref{sec:simulation_methods}.

Fig.~\ref{fig:example_phenomenological_phasediagram} shows the threshold surface for the 4-star (blue line) and 6-ring (orange line) fusion networks. If the combination of erasure probability $p_{\rm erasure}$ and error probability $p_{\rm error}$ per measurement lies inside the threshold surface, the errors are in the \textit{correctable region} of the corresponding fusion network. In the correctable region the probability of logical errors is suppressed exponentially in the size of the network. The correctable region of the 4-star network is contained in the correctable region of the 6-ring network so any value of ($p_{\rm erasure}$, $p_{\rm error}$) that can be corrected by the 4-star network can also be corrected by the 6-ring network. 
The marginal $p_{\rm erasure}$ threshold for the 4-star network is $6.90\%$, while it is $11.98\%$ for the 6-ring network. 
The marginal $p_{\rm error}$ threshold for the 6-ring network is $1.07\%$, compared to $0.75\%$ for the 4-star network.

\subsubsection{Linear optical error model}
\label{subsec:linear_optic_FTFN_example}

\begin{figure}
\includegraphics[width=\columnwidth]{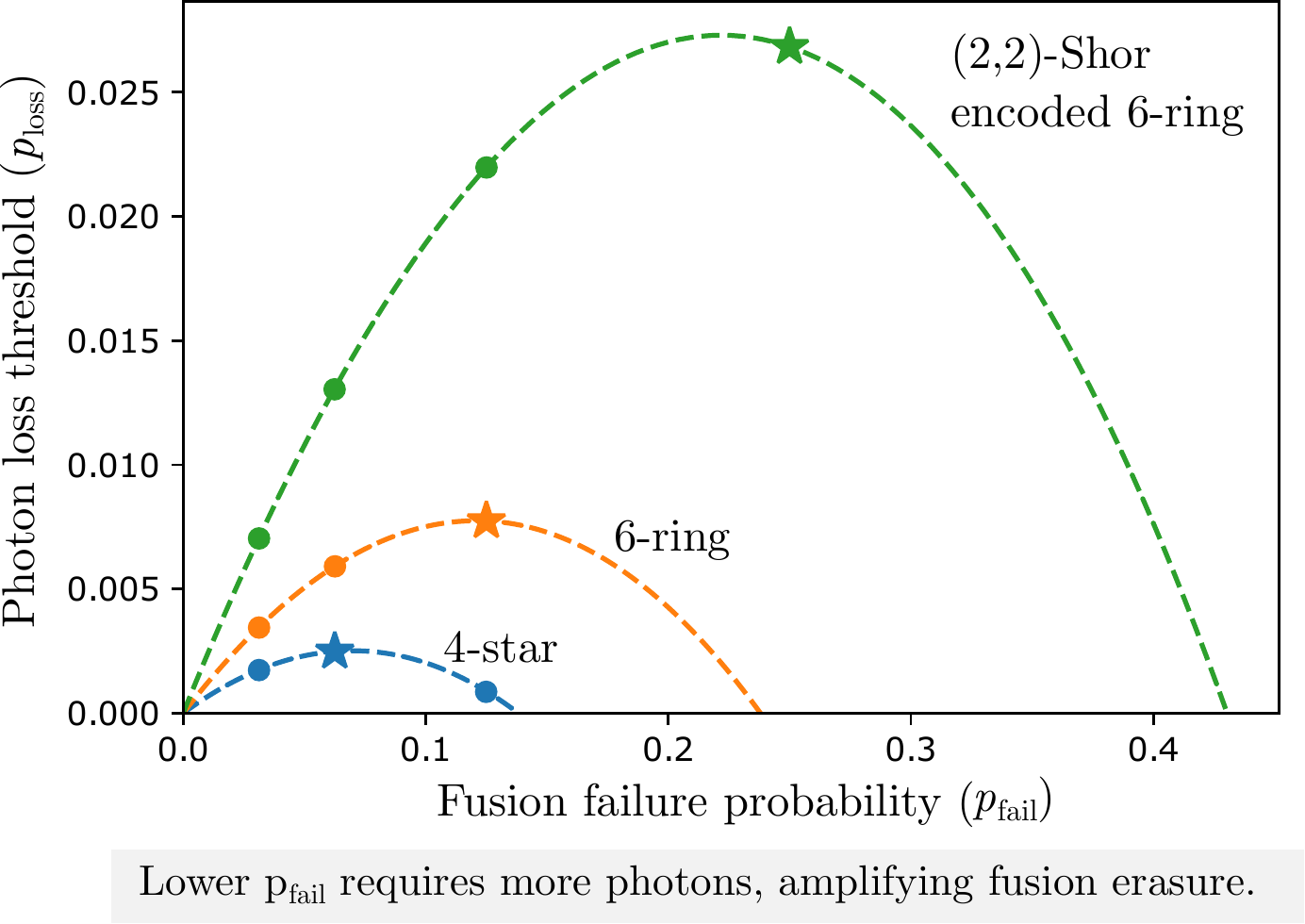}
\caption{Photon loss threshold for the three fusion networks: 4-star (blue) and 6-ring (orange) and (2,2)-Shor encoded 6-ring (green). The threshold is calculated under the linear optical error model with the same photon loss probability $p_{\rm{loss}}$ applied to every photon in the protocol. We consider a physical model for fusion failure where $p_{\rm{fail}} = 1/2^n$ can be achieved by boosting a fusion with $2^n - 2$ additional photons. Since more photons are required for these lower fusion failure rates, the effect of loss in this regime is amplified, with a probability $(1-p_{\rm loss})^{2^n}$ of no photon in the fusion being lost. Because of this the protocols demonstrate an optimal performance at some intermediate value of $p_{\rm fail}$. The markers represent the values of $p_{\rm{fail}}$ that can be achieved with fusions presented in \cite{grice2011arbitrarily} and the stars represent the optimum levels of boosting for the different schemes.
The green curve corresponds to the 6-ring fusion network with qubits encoded in a (2,2)-Shor code. 
The details of the encoding and measurement scheme, and the error model used to evaluate these curves is explained in appendix~\ref{app:fusion_error_model}. 
}
\centering
\label{fig:example_lossfail_phasediagram}
\end{figure}

We now examine the performance of these fusion networks under an error model motivated by linear optics, which includes the effects of fusion failure and photon loss. Every fusion is a linear optical Bell fusion between dual-rail qubit photonic resource states as described in Sections~\ref{sec:fusion} and~\ref{sec:resource-states}. 

Tolerance to photon loss in this model differs from erasure tolerance. Erasure can arise in multiple ways. Firstly in every fusion there may be multiple photons involved, each of which can experience loss, the more photons there are the greater the chance that one will be lost and lead to an erasure. In addition to loss fusion failure can also lead to an erasure. In the absence of loss, every fusion performs the intended measurements with probability $1-p_{\rm{fail}}$ and performs separable single qubit measurements with probability $p_{\rm{fail}}$. As described in section~\ref{sec:fusion}, this failure case can be treated as one of the two intended fusion measurements being erased. Here we choose the linear optical fusion circuit for each fusion randomly so that both fusion measurements have an equal probability of being erased due to fusion failure. 

To model fusion failure we consider the family of fusion boosting protocols in~\cite{grice2011arbitrarily}. For unboosted fusion $p_{\rm{fail}} = 1/2$ and no ancilliary photons are required. If the fusion is boosted with a Bell pair, $p_{\rm{fail}}$ = 1/4 and there are two ancilliary photons in the fusion. In general, \cite{grice2011arbitrarily} shows that $p_{\rm{fail}} = 1/2^n$ can be achieved by boosting a fusion with $2^n - 2$ additional photons. For a fusion on $N$ photons the probability that no photon in the fusion is lost is $\eta^N$, where $\eta = 1-p_{\rm loss}$. If any photon in a fusion is lost, fewer than expected photons are detected and both fusion outcomes are erased. We therefore use a model of fusion erasure that captures the tradeoff between $p_{\rm{fail}}$ and the probability of losing a photon. The probability that no photon in the fusion is lost is $\eta^{1/p_{\rm{fail}}}$. Hence, with probability $1 - \eta^{1/p_{\rm{fail}}}$, a fusion is erased. In this error model every individual physical fusion measurement in the network has an erasure probability of $p_0 = 1 - (1 - p_{\rm{fail}}/2)\eta^{1/p_{\rm{fail}}}$, which we explain in detail in Appendix~\ref{app:fusion_error_model}.

We numerically model three fusion networks under this model of fusion and photon loss. The 4-star network, the 6-ring network, and an encoded 6-ring network. In the encoded version of the fusion network every qubit in the resource state of the 6-ring network is replaced by a qubit encoded in the (2,2)-Shor code as depicted in Fig. \ref{fig:resource_state_6ring}. 
With this replacement, fusions between the (unencoded) qubits in a resource state are replaced by encoded fusions between the encoded qubits in resource states, which are composed of pairwise fusions between the physical qubits that constitute an encoded qubit.
The (2,2)-Shor code refers to a four qubit $[[4,1,2]]$ quantum code which can be obtained by concatenating repetition codes for $X$ and $Z$ observables. (2,2) refers to the sizes of the two repetition codes.
Depending on the order of concatenation, the resulting code space will be described by the code stabilizers $\langle XXXX, ZZII, IIZZ \rangle$ or $\langle ZZZZ, XXII, IIXX\rangle$. 
For simplicity, we assume that this choice is taken uniformly at random for every encoded fusion. 
An encoded fusion on two qubits $A$ and $B$ attempts to measure the input encoded qubits in the encoded Bell basis $\overline{X_A}\overline{X_B}, \overline{Z_A}\overline{Z_B}$ where $\overline{X}$ and $\overline{Z}$ respectively denote encoded $X$ and $Z$ operators for the local (2,2)-Shor code that are explicitly defined in appendix \ref{subapp:encoded_fusion_erasure_prob}. The local stabilizers in an encoded qubit allow the encoded fusion measurement outcome to be reconstructed in multiple ways which suppresses the erasure rate for these encoded fusions. 
In Appendix \ref{subapp:encoded_fusion_erasure_prob}, we explicitly calculate the erasure probability for measurements in an encoded fusion. 
If the erasure probability of unencoded fusion measurements is $p_0$, the probability of encoded fusion measurement erasure is $$p_{\rm enc} = \frac{[1 - (1-p_0)^2]^2 + 1 - (1-p_0^2)^2}{2}.$$
For $p_0 < 0.5$, $p_{\rm enc} < p_0$ i.e. the encoding suppresses the erasure probability.

There are three levels of encoding which we are separately modeling:
\begin{itemize}
    \item At the lowest level is the linear optical qubit encoding of representing each physical qubit as a dual-rail qubit, where loss and multi-photon errors take the state outside the computational subspace.
    \item A local encoding (the (2,2)-Shor code) is used in resource states to achieve an encoded fusion which is less susceptible to fusion failure and photon loss. 
    \item At the highest level we construct a fusion network which defines a topologically-protected logical qubit.
\end{itemize}

To evaluate the fusion networks in the linear optical error model we do not need to perform numerical simulations, but we can instead perform a mapping between $p_{\rm fail}$ and $p_{\rm loss}$ and the erasure parameter of our hardware-agnostic fusion error model: $p_{\rm erasure}$, and use the simulated threshold values from Figure~\ref{fig:example_phenomenological_phasediagram}. 
In Fig.~\ref{fig:example_lossfail_phasediagram}, we plot the threshold in photon loss of the fusion networks described above as a function of the fusion failure probability $p_{\rm{fail}}$. Although a low value of $p_{\rm{fail}}$, which is achieved by boosting the fusion, reduces erasure due to fusion failure, our model penalizes high levels of boosting by accounting for the loss on the increase number of boosting photons needed to achieve these low failure rates. As a result, there is an optimum value of $p_{\rm{fail}}$ for every fusion network which corresponds to an optimum level of boosting.

The blue and orange lines in Figure~\ref{fig:example_lossfail_phasediagram} show the threshold behavior for the 4-star and 6-ring fusion networks respectively. The failure thresholds for these networks (in the absence of photon loss) is below $25\%$, which means that simple boosted fusion is not sufficient for fault tolerance. The markers represent the values of $p_{\rm{fail}}$ that can be achieved with fusions presented in \cite{grice2011arbitrarily} and the stars represent the optimum levels of boosting. With the (2,2)-Shor encoding, the 6-ring fusion network provides a significantly larger marginal failure threshold of $43.2\%$. With the $25\%$ failure probability achieved with fusions boosted with a Bell pair, we have a loss tolerance of $2.7\%$ per photon. In other words, by boosting fusions with a Bell pair, the fusion network can be in the correctable region even when the probability of at least one photon being lost in a fusion is $10.4\%$.

\section{Quantum computation with fault-tolerant fusion networks}
\label{sec:quantum-computation}

The previous sections have described how to create a fault tolerant \emph{bulk} in FBQC - which behaves as the fabric of topological quantum computation. Creating the bulk is the most critical component of the architecture, as it is this that determines the error correction threshold. But to implement fault-tolerant \emph{computation}, additional features are needed. We now turn to the question of how this bulk can be used to implement fault-tolerant logic, and the implications for classical processing and physical architecture. Here we provide a brief overview of logic in FBQC and architectural considerations, both of which are topics in their own right and which are explored in much greater depth in related publications~\cite{interleaving, logic_blocks}.

\subsection{Logical Gates}

In order to perform fault-tolerant logic, we need to add to our toolkit the ability to create \emph{topological features} in addition to the bulk. 
There are different approaches that can be used to create a fault-tolerant Clifford gate set. 
Boundaries can be used to create punctures which can be braided to perform gates via code deformation~\cite{Briegel_2009,bombin2009quantum}.
Boundaries can be used to create patches on which lattice surgery can be performed~\cite{horsman2012surface}.
Logical qubits can alternatively be encoded in defects and twists~\cite{bombin2010topological,brown2020universal,webster2020fault}. All of these approaches to logic are compatible with FBQC. 
The necessary topological features can be created by modifying fusion measurements in certain locations, or adding single qubit measurements in an appropriate configuration.
Here we give a simple example of how to create two types of boundaries, which is sufficient to enable the encoding and manipulation of logical qubits in punctures or patches. Other topological features are addressed in~\cite{logic_blocks}.

\subsubsection{Boundary creation}
These two boundary types we create correspond to rough and smooth boundaries in the surface code picture \cite{bravyi1998quantum}, but in FBQC it is more natural to refer to them as primal and dual boundaries, according to whether they are able to match excitations in the primal/dual syndrome graphs respectively. A primal boundary corresponds to a rough boundary in the primal syndrome graph, and to a smooth boundary in the dual syndrome graph.

\begin{figure*}
    \centering
    \includegraphics[width=1.7\columnwidth]{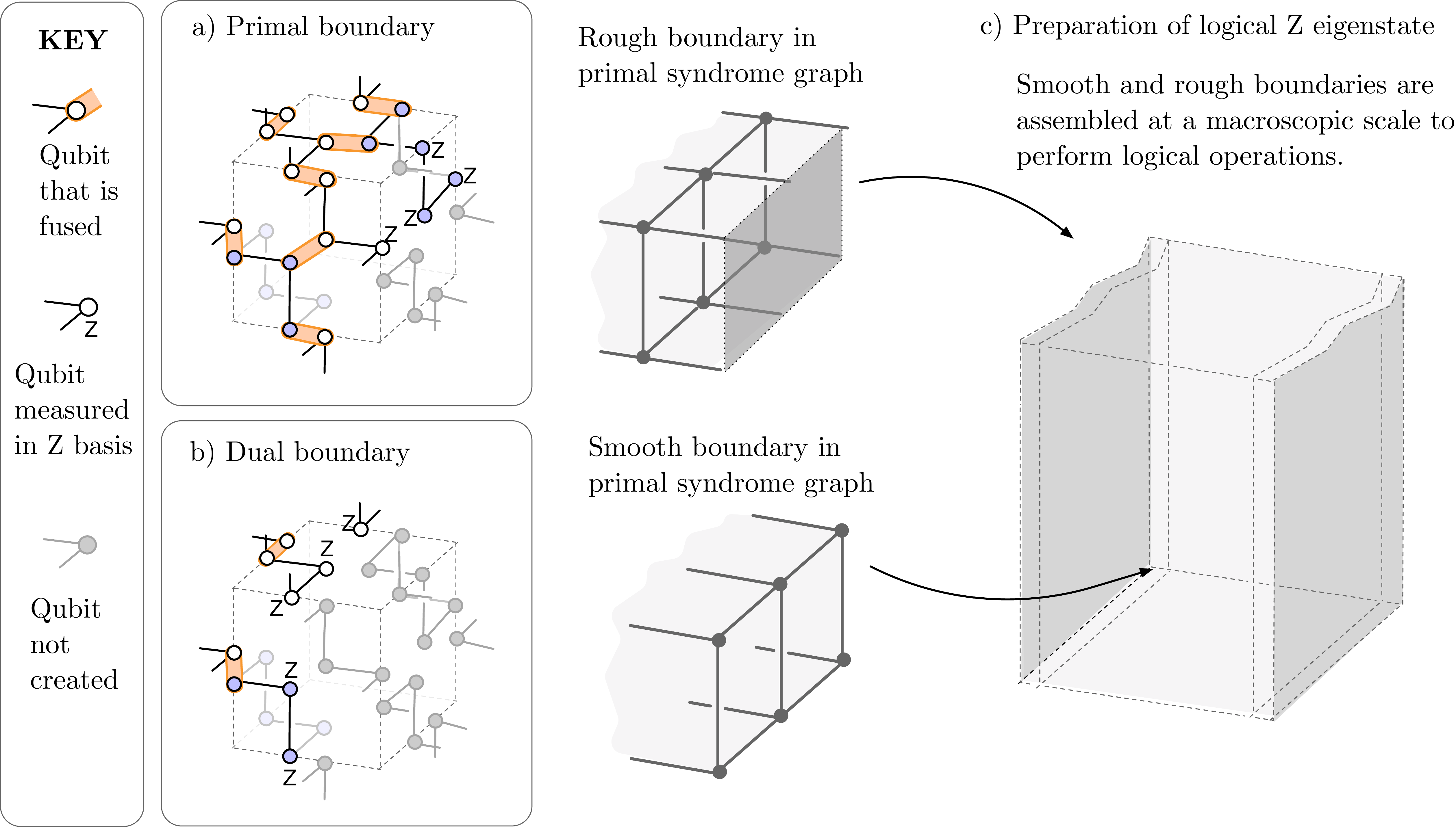}
    \caption{A scheme for creating primal and dual boundaries that can be used to modify the bulk to perform quantum computation. a) and b) show modified unit cells of a fusion network that can generate a primal and dual boundary respectively. In each, the fusion network is made up of the same configuration of resource states as in the bulk (see Figure~\ref{fig:kagome6_layout}), but where some subset of the fusion measurements have been replaced with single qubit Z measurements, and some subset of the resource states are entirely removed (indicated by greyed out circles). If at the boundary a resource state has no remaining entangling operations connecting it into the bulk, then it does not need to be created. All the remaining fusions (shown by orange ovals) are a projective measurement on $XX$, and $ZZ$. The effect of this modified network is to truncate the bulk either at (a) a slice halfway through the cell or (b) at the edge of the cell. A primal (dual) boundary terminates a primal (dual) logical membrane, and therefore has a rough boundary in the primal (dual) syndrome graph. The primal (dual) boundary has a smooth boundary in the dual (primal) syndrome graph. (c) Shows an example of hows these unit cells can be composed to create macroscopic boundary conditions, enabling fault-tolerant logic. This shows logical block that corresponds to fault-tolerant preparation of a dual membrane.}
    \label{fig:creating_boundaries}
\end{figure*}

Figure~\ref{fig:creating_boundaries} illustrates a simple example of how primal and dual boundaries can be created by measuring certain qubits of resource states in the $Z$ basis. Figure~\ref{fig:creating_boundaries}(a) shows the creation of a boundary, where the layer of qubits at the boundary of the unit cell are either measured in the Z basis, or simply never created. 
Figure~\ref{fig:creating_boundaries}(b) shows a similar protocol for creating a smooth boundary. 
The case in which boundaries are parallel to planes defined by pairs of the unit cell vectors are particularly simple.
For such boundaries, the only distinction between the primal and dual case is a displacement by half a unit vector in the perpendicular direction. The effect of this measurement pattern is to terminate the bulk, creating boundaries which can then be used as a feature to encode and manipulate logical qubits. Figure~\ref{fig:creating_boundaries}(c) shows an example of how these boundaries can be macroscopically assembled to fault tolerantly prepare the state $\ket{0}$ (or $\ket{1}$)  in a patch encoded logical qubit.

\subsubsection{Pauli frame tracking}
In FBQC, logical states have a direct physical counterpart only up to a Pauli correction, which is tracked in classical logic through the so called {\it Pauli frame}.
The use of a Pauli frame is necessary due to the intrinsic randomness which is introduced by teleportations carried out by Bell measurements.
For instance, at the logical level, the same component which prepares a $\ket{0}$ state in Figure~\ref{fig:creating_boundaries} will also represent the preparation of a $\ket{1}\equiv X\ket{0}$ state.
In general, this means \cite{Chamberland_2018, Bertels17} that a state $\ket{\psi}$ may be physically represented by a different state $P\ket{\psi}$ for some tracked Pauli correction operator $P$.

When relying on Pauli frame tracking, a generic $n$ qubit state can be represented by any of $4^n$ possible physical quantum states together with $2n$ classical bits which describe frame. 
The use of stabilizer codes which protect the logical information practically halves the number of bits required to describe the frame.
The key property of this technique is that most of the computation which can be described by Clifford operations can be executed independently of the classical tracking information.
The classical Pauli frame data only influences the quantum operations performed at the logical level, in cases such as magic state injection and distillation.
This allows classical Pauli frame processing to occur at a logical clock rate rather than at a potentially much faster physical fusion clock rate.
In appendix \ref{sec:pauli_frame}, we explain Pauli frame tracking for fault-tolerant FBQC, explaining why this technique only imposes minimal quantum and classical processing requirements.

\subsubsection{Universal logic}
To achieve a universal gate set, the Clifford gates must be supplemented with state injection, which combined with magic state distillation protocols can be used to implement $T$ gates, or other small angle rotation gates. 
Magic state injection can be implemented in FBQC by performing a modified fusion operation, by making a single qubit $\frac{\pi}{8}$ measurement, or by replacing a resource state with a special `magic' resource state.

\subsection{Decoding and other classical processing}

In FBQC, as in other approaches to fault-tolerant quantum computation, classical error-correction protocols are in charge  of extracting reliable logical measurement information from the unreliable and noisy physical measurement outcomes. In FBQC it is helpful to view the decoding outcomes as \emph{logical Pauli frame information}. Keeping track of this logical Pauli frame is necessary to interpret future measurement outcomes. 

This logical Pauli frame produces time-sensitive information when logical level feed-forward is required. That is, when a logical measurement outcome is used to decide on a future logical gate it is necessary to have the relevant Pauli frame information available. One example of this is for the realization of $T$-gates via magic state injection where a $S$ or $S^{\dagger}$ is applied conditioned on a logical measurement outcome.

One widely discussed challenge of decoding is that it must be performed live during quantum computation. However, it is a crucial feature that this feed-forward operation happens at the \emph{logical timescale}, and decoding outcomes are not needed at the fusion (or physical qubit) timescale. If decoding is slower than the logical clock rate then \emph{buffering}  or \emph{ancillary logical qubits}~\cite{litinski2019game} can be used to allow the computation to `wait' for the decoding outcomes. We discuss this further in Appendix~\ref{app:decoding}. It is worth emphasizing that these are tools used at the logical level, and it is never necessary to modify any physical operation. Fusions can always proceed without decoding outcomes.  An important implication of this is that a slow decoder does not impact threshold. Nevertheless fast decoders are desirable to reduce unnecessary overhead.

\subsection{Physical architecture}
\label{sec:physical_arch}

\begin{figure*}
    \centering
    \includegraphics[width=1.5\columnwidth]{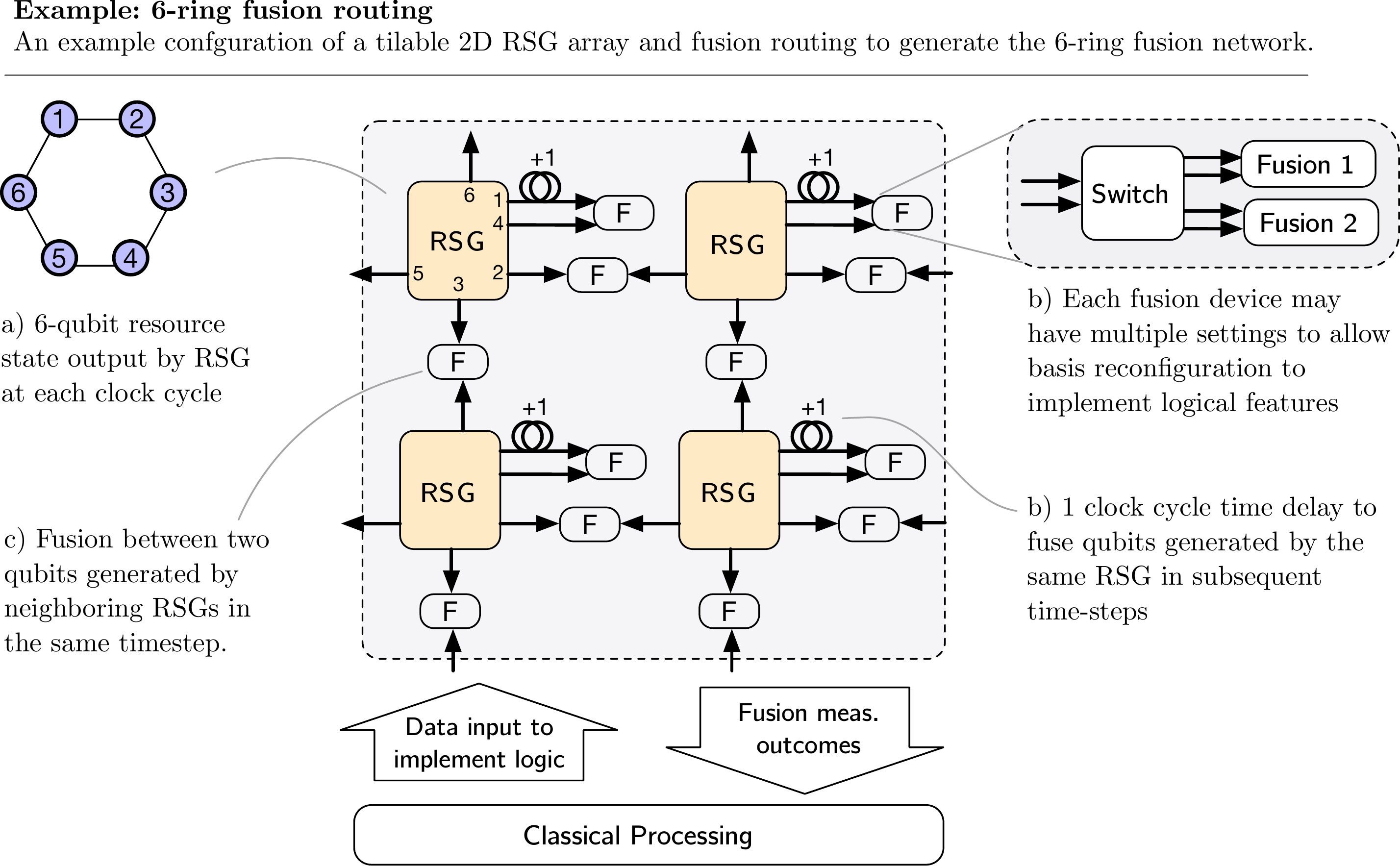}
    \caption{Example of a physical layout of resource state generators and fusion routing that can be used to create the 6-ring fusion network. a) Four RSGs are shown, each producing a 6-ring state in each clock cycle. These are arranged in a tileable configuration. Qubits from each RSG are routed to 2-qubit fusion devices. b) Each fusion device can include a switch that can reconfigure between multiple fusion settings to implement logical gates. Each RSG outputs 6 qubits per clock cycle. c) Four qubits from each state immediately enter a fusion device in one of the four spatial directions: North, South, East or West. This generates entanglement between states created at neighboring sites in the same time step. d) The two remaining qubits from each state are used to generate entanglement between states produced at the same physical location, but in different time steps. To achieve this, one qubit passes through a 1 clock cycle delay, so that it arrives at the fusion device coincidentally with a state produced in the following clock cycle. Fusion measurement outcomes are output from the system as classical data. In the bulk no data input is required, but classical control is needed at certain locations to reconfigure fusion devices to perform logical gates.}
    \label{fig:fusion_routing}
\end{figure*}

Fusion networks have no intrinsic notion of time or space. The fusion network does not specify the ordering of fusion measurements, nor is it necessary that all of the resource states exist simultaneously. The same fusion network could be implemented by producing all the resource states at the same time or by staggering the resource state generation such that only some portion of the fusion network is `alive' at a given moment in time. Re-introducing physical space, and time-ordering is an architectural design tool that can be used based on the available hardware. At the center of an architecture is the mechanism for generating resource states, which will be generated at a certain spatial location and a certain time. It is natural to consider the notion of \emph{resource state generators} (RSGs), which correspond to physical devices that produce resource states at a certain clock speed. This picture is particularly relevant for photonic architectures. 
We can then consider the lifetime of a qubit, which is created in a resource state generator, passed into the fusion network router, which routes qubits to the right fusion location. The qubit is then destructively measured in fusion. This very limited qubit lifetime is a strength of FBQC, particularly in photonic architecture where optical loss is the dominant source of physical error. 
In Figure~\ref{fig:fusion_routing}, we show an example configuration of a tilable 2D array of RSGs connected to fusions by a fusion network router to implement the 6-ring example fusion network introduced in section \ref{sec:example_6ring}. This example demonstrates several important features of the physical implementation of a scheme for FBQC: 

\begin{enumerate}
\item {\bf The operational depth is extremely low.} Each qubit is initialized and entangled into a resource state, and then immediately measured in fusion. The qubit's lifetime is independent of the size of the code, or the computation. 
\item {\bf Resource state generators can be repeatedly used to create a large fusion network.} A fusion network contains many resource states, but they do not all need to co-exist simultaneously. An RSG that produces a state in each clock cycle can be re-used repeatedly. One natural way to time-order the creation of a large fusion network is to divide it into `time slices'. A 3D fusion network is divided into 2D layers, with one layer created at each time step, and fused with the previous layer. This time ordering allows a 3D network to be generated with a 2D array of resource state generators. 
\item {\bf The fusion routing can be fixed, such that rerouting between each clock cycle is not required.} A good design principle in a fusion routing is to minimize the need for switching between clock cycles. This reduces loss and error from switching, and minimizes the need to input classical control signals. In this example layout, every resource state generated at a given location goes to the same fusion devices. This means that the connections of the device are fixed, and no switching is needed to generate the bulk. 
\item {\bf Logic can be implemented by modifying fusion measurements.} To allow logic to be implemented (at least) a subset of the fusion devices must be reconfigurable, as indicated in Figure~\ref{fig:fusion_routing}b). Boundaries or other topological features in the bulk are implemented by changing the measurement basis of the fusion, or switching to single qubit measurements. 
\end{enumerate}

The example in Figure~\ref{fig:fusion_routing} represents a simple physical architecture. With photonic qubits in particular there is a lot of flexibility in how to construct such an architecture. 
Depending on how long resource states can wait in memory before suffering too much decoherence or loss, more extreme time-ordering approaches can be taken. With photonic resource states the ideas of \emph{interleaving}, described in~\cite{interleaving}, can be applied to use a single RSG to create an entire block of a fusion network by creating the network one resource state at a time. As long as the time ordering of state creation is compatible with higher level feedforward constraints at the logical level, any time ordering is possible. 
This example includes only local connectivity, which is not a requirement for photonic qubits. Long-range connectivity can allow the creation of non periodic boundary conditions, other topologies, or codes embedded in non-euclidean spaces. Higher dimensional fusion networks can also be created by combining time ordering structures with appropriate optical connectivity.

\subsection{Errors in FBQC}
\label{subsec:FBQC_errors}

The thresholds presented in Section~\ref{sec:example} are based on simple error models. In a physical implementation there will many physical sources of imperfection that contribute to the total erasure and measurement error rates, as well as the specific structure of those errors. A full analysis requires a detailed system architecture, and depends strongly on the specifics of physical error models. However the models we present here can still be used to get a meaningful insight into realistic performance. Since our model does not fix the ratio of Pauli and erasure errors, and since some correlated error structure is already present in the error model it is often possible that an exact or approximate analytical mapping can be made from a more detailed circuit-level error model to the numerical results we presented in Sec.~\ref{sec:example}. 
When it comes to specific structure in the errors, error bias and correlations impact the threshold, and time ordering of operations can spread errors. However, there is reason to believe that the impact of these can be limited in FBQC. In particular: 
  
\begin{enumerate}

\item \textbf{FBQC accounts for the structure of errors due to the creation of long range entanglement}. As we build up large scale entanglement from low-weight physical operations, errors in resource states will lead to fusion measurement errors. The way errors propagate from resource state generation through fusion measurements is captured in the syndrome graph.

\item \textbf{Resource state and fusion errors are intrinsically local}. The construction of FBQC limits how far errors can potentially spread. Assuming they are created in physically separate systems we would expect correlations to exist only within a resource state and not between resource states (prior to fusion). This expectation is particularly strong with linear optics, where photons at different locations cannot become `accidentally' entangled with one another since they do not interact. Furthermore, each qubit in the protocol has a short finite lifetime, limiting the potential for the spread of errors in its neighborhood.

\item \textbf{Correlations within a fusion can only improve performance}. A likely place for correlated errors to appear is between the two measurement outcomes of a fusion operation. Our model treats these errors as uncorrelated. Since we decode primal and dual syndrome graphs separately here, if fusion errors were correlated it would make no difference to our thresholds. If that information were to be accounted for in decoding it could only improve the performance.
\end{enumerate}

Finally, when considering computation we need to account for the fact that logic gates are performed via creating topological features, such as boundaries or twists, which need different physical operations. We would therefore expect these to have different error models at those locations. It is, however, the case that in topological fault tolerance the bulk determines the threshold. The topological features used to implement logic are 2- or 1-dimensional objects. Our numerical results should therefore correctly indicate the threshold of fault-tolerant logic, although logic gates may have a different below threshold scaling behavior.

\section{Discussion}
\label{sec:discussion}

FBQC is a model for universal quantum computation. At the level of fault-tolerant logical gates it permits the same operations as circuit-based quantum computing (CBQC) or measurement-based quantum computing (MBQC)~\cite{Briegel_2009}. 
But significant differences emerge when looking at the physical processes used to implement the logical gates: in the resources needed with the computational protocol, in the operational depth, in the required connectivity of physical qubits, in the processing of classical information and in the emergence, propagation and impact of errors.

Both FBQC and MBQC are models based on resource states and measurements. The key distinction is in the nature of the resources. MBQC requires a large entangled cluster state, of a size that scales up with the computation being performed. FBQC, on the other hand, requires resource states of a constant size where the number of resource states needed increases for a larger quantum computation.
The distinction is also clear in the type of measurements used. MBQC uses single qubit measurements to perform computation, and previous work that propose linear optical quantum computing (LOQC) architectures to achieve fault-tolerant MBQC~\cite{Knill_2004,knill2004fault,Dawson_Haselgrove_Nielsen_2006,Herrera-Marti2010,li2015resource,auger2018fault,herr2018local} did so by first creating the large cluster state resource out of finite size operations, and following this with computational (single-qubit) measurements. No such separation exists in an FBQC protocol, where destructive multi-qubit projective measurements\cite{leung2004quantum,verstraete2004renormalization,nielsen2003quantum}, such as fusion gates\cite{browne_rudolph}, integrate the entangling measurements needed to create long-range entanglement with the measurements that implement fault tolerance and computation. Although not required, in some variations of FBQC the protocol may also  contain a small number of single qubit measurements, for example to create topological features~\cite{logic_blocks}. 

When it comes to fault-tolerant computation with linear optics, more distinctions arise at the architectural level. LOQC has a long history, with the earliest proposals~\cite{KLM, nielsen2004optical, browne_rudolph} relying on extremely large gate teleportation or repeat-until-success strategies to handle probabilistic gates, and as such, requiring quantum memories.  More recent architecture proposals~\cite{kieling2007percolation,GSBR} replaced the need for memory with that of short fixed delays, however, the schemes for fault tolerance were still based on building a large entangled cluster state, and then making single qubit measurements on that state to implement fault tolerance and computation via MBQC. These schemes have a low \emph{constant depth}, meaning that each photon sees a small fixed number of components during its lifetime, regardless of the size of the computation. The highest performing schemes were based on percolation methods\cite{kieling2007percolation, GSBR} to handle probabilistic fusion. Other schemes used branched resource states to add redundancy~\cite{li2015resource}, which were able to tolerate probabilistic fusion at the expense of reducing the threshold against loss and Pauli error. 

The FBQC schemes we introduced in section~\ref{sec:example} use constant sized resources in an architecture with a constant depth, but offer a significant threshold improvement compared to the best results in the literature.
Beyond error tolerance, FBQC offers a crucial advantage in architectural viability compared to these previous schemes, where classical processing and feedforward was required to happen during the lifetime of a photon. This classical processing is often complex~\cite{herr2018local}, and the need to perform a global computation of this kind while a photon waits in a delay line would place extraordinary performance requirements on photonic delays. FBQC removes this requirement and makes the fault tolerance threshold independent of the timescale of classical feedforward once resource states are available. As discussed in Section~\ref{sec:quantum-computation}, feedforward is still needed, as is the case in any quantum computing architecture, but in FBQC this requirement is only at the logical level, with a timescale completely separated from physical operations. 

We have presented an example of FBQC for the example of linear optical quantum computing with dual-rail qubits, however these tools can be applied across many hardware platforms to construct architectures that are intrinsically modular and have a low operational depth on each physical qubit between initialization and measurement. 
A low depth of operations limits the propagation of errors, and allows leakage to be identified and corrected. In addition, minimizing the operational depth is likely valuable wherever measurement operations are simpler and less noisy to perform than entangling gate operations. These benefits may apply to both photonic and non-photonic architectures. Modularity is also broadly desirable, as it is a key aspect ensuring an architecture for quantum computing is reliable and manufacturable. The functional blocks of FBQC: resource state generation, fusion network and fusion operations, must be compatible with each other, but the physical implementation of each block can differ as long as they are compatible. This opens the door to supporting applications in \emph{hybrid} quantum systems. For example, photonic fusion operations in a fiber-based fusion network could be integrated with matter-based resource state generators producing photonic entangled states. Modularity of FBQC in optical quantum computing, coupled with the flexibility of optical connections can enable many possible fault tolerance schemes beyond the 3D topological schemes presented here~\cite{interleaving}.

FBQC is a general framework for computation that allows us to link together physical operations and topological fault tolerance, yielding low depth and modular architectures. This is a tool that can be used to design and optimize architectures for fault-tolerant quantum computing that are modular, and have a low operational depth. We expect this framework, with its ability to tightly link physical errors with their effect on quantum error correction, will allow performance improvements in systems that are fundamentally based on resource state generation and projective measurements. As technology moves ever closer to realizing these systems, having such a theoretical framework will be an important tool for engineering hardware design to achieve large scale fault-tolerant quantum computation.

\section*{Acknowledgements}
The authors would like to thank 
Nikolas Breuckmann,
Jacob Bulmer,
Axel Dahlberg,
Andrew Doherty,
Megan Durney,
Nicholas Harrigan,
Isaac Kim,
Daniel Litinski,
Ye-hua Liu,
Kiran Mathew,
Ryan Mishmash,
Sam Morley-Short,
Andrea Olivo,
Sam Pallister,
William Pol, 
Sam Roberts,
Karthik Seetharam, 
Jake Smith,
Jordan Sullivan, 
Andrzej P\'erez Veitia 
and all our colleagues at PsiQuantum for useful discussions. Terry Rudolph is on leave from Imperial College London, whose support he gratefully acknowledges.

\appendix

\section{4-star and 6-ring fusion networks}
\label{app:formal_fusion_network_definitions}
Here we give an explicit geometric definition of the fusion networks introduced in Section~\ref{sec:example} to accompany Figure~\ref{fig:star_layout} and Figure~\ref{fig:kagome6_layout}. The geometry is merely a tool for defining the connectivity of the fusion network, and the physical positioning on the qubits and fusions has no intrinsic meaning. 

\paragraph{4-star fusion network}

Each cubic unit cell contains 6 resource states. Three are positioned on unit cell edges at locations (1/2, 0,0), (0,1/2,0) and (0,0,1/2). Three are positioned on unit cell faces at locations (1/2,1/2,0), (1/2,0,1/2) and (0,1/2,1/2). Each resource state has four qubits, which are all equivalent. In addition the unit cell contains 12 two-qubit fusion operations. Every resource state positioned at a face undergoes a fusion with the resource state located at the four edges making up the boundary of the face. All fusions are a stabilizer measurement of the operators $XZ$ and $ZX$ on the two measured qubits. 
    
\paragraph{6-ring fusion network}

Resource states are distributed according to a body-centered cubic lattice, centered at locations {(0,0,0), (1/2,1/2,1/2)} in the unit cell.
Additionally, we associate individual coordinates to the qubits of the resource states according to the mapping, where these positions are relative to the resource state position: 

\begin{align*}
& 1: (1/4,0,0), \\
& 2: (1/4,1/4,0), \\
& 3:(0, 1/4, 0), \\
& 4:(0, 1/4, 1/4), \\
& 5:(0,0,1/4), \\
& 6:(1/4,0, 1/4). 
\end{align*}

There are two qubits located at each edge, which are connected by a fusion, and two qubits located at each face, which are also connected by a fusion.

\section{Linear optical fusion error model}
\label{app:fusion_error_model}

A fusion attempts to measure two input qubits $1$ and $2$ in the Bell basis $X_1X_2$, $Z_1Z_2$. However linear optical fusion on two dual-rail qubits does not always perform these measurements because of the inherent non-determinism of dual-rail entanglement generation with linear optics and the presence of photon loss. In the example in Section~\ref{sec:example}, we primarily consider linear optical Bell fusion where every photon, including the photons used in boosting, is lost with the same probability $p_{\rm{loss}}$ and $\eta := 1-p_{\rm{loss}}$. 
In the linear optical error model, we assume that the probability of a fusion having no lost photon is $\eta^{1/p_{\rm{fail}}}$ where $p_{\rm{fail}}$ is the probability of the fusion failing. This model accounts for the fact that boosting to obtain lower fusion failure probabilities involves using a larger number of photons as ancillas in the fusion, which implies a higher risk of fusion erasure due to photon loss. If the fusion is unboosted, $p_{\rm{fail}} = 1/2$, there are only two photons in the fusion and the probability that no photon in the fusion is lost is $\eta^2$. If the fusion is boosted with a Bell pair~\cite{grice2011arbitrarily}, $p_{\rm{fail}}$ = 1/4, there are four photons in the fusion (two input photons and two photons from the ancilla Bell pair) and the probability that no photon in the fusion is lost is $\eta^4$. In general, \cite{grice2011arbitrarily} shows that $p_{\rm{fail}} = 1/2^n$ can be achieved by boosting a fusion with $2^n - 2$ additional photons, resulting in a probability $\eta^{2^n}$ of no photon in the fusion being lost. If any photon in a fusion is lost, fewer than expected photons are detected and both fusion outcomes from the fusion are erased. 

When all photons are detected in the fusion, a fusion fails with probability $p_{\rm{fail}}$, and instead of the intended measurements of $X_1X_2$ and $Z_1Z_2$, the fusion performs separable single qubit measurements. Depending on the linear optical circuit used to perform the fusion, the fusion can measure a pair of single qubit stabilizer measurement (e.g. $X_1$ and $X_2$ or $Z_1$ and $Z_2$) when it fails\footnote{It is simple to modify linear optical circuits to choose the failure basis using appropriate single qubit gates, which are easy to implement in linear optics, before a fusion. For instance a fusion that measures $Z_1$, $Z_2$ on failure can be made to fail by measuring $X_1$, $X_2$ instead by placing a Hadamard gate before both input qubits.}. By taking the product of these two single qubit measurements we can reconstruct the two qubit measurement, therefore this event can be interpreted as a successful fusion with an erasure of one of the measurement outcomes. For example, if the intended fusion measurements were $X_1X_2$ and $Z_1Z_2$ and, upon fusion failure, we obtain single qubit measurement outcomes $X_1$ and $X_2$, we can treat this case as a successful fusion with an erased $Z_1Z_2$ measurement outcome

In this paper, the circuits used to implement fusion are randomized so that with $50\%$ probability, the fusion measures $X_1,X_2$ on failure, and with $50\%$ probability, $Z_1,Z_2$ are measured on failure. The probability that no photon in the fusion is lost is $\eta^{1/p_{\rm{fail}}}$ and the erasure probability for a fusion measurement in the absence of loss is $p_{\rm{fail}}/2$ in this randomized model. Consequently, the erasure probability for both the $X_1X_2$ and $Z_1Z_2$ measurements coming from the fusion is $p_{\rm{fail}}\eta^{1/p_{\rm{fail}}}/2$.  The probability of measuring both $X_1X_2$ and $Z_1Z_2$ is $(1-p_{\rm{fail}})\eta^{1/p_{\rm{fail}}}$. With this randomization of the failure basis, the marginal probability of erasure for individual measurements coming from the fusion is 

\begin{equation}
p_0 
= 1 - (1 - p_{\rm{fail}}/2)\eta^{1/p_{\rm{fail}}},
\label{eq:p0_fusion_meas_erase_prob}
\end{equation}

\noindent which we call the physical fusion measurement erasure probability. Since the two measurements from a fusion go to different syndrome graphs (primal and dual) that are decoded separately, evaluating the fault-tolerance threshold only requires this marginal erasure probability. Table \ref{table:fusion_outcomes} summarizes the probability of obtaining different measurements in a fusion boosted by a Bell pair with $p_{\rm{fail}} = 1/4$. 

\begin{table}[h!]
\centering
\begin{tabular}{ |p{1.5cm}||p{1cm}|p{1cm}|p{1cm}|p{1cm}| } 
 \hline
 Fusion outcome & success & failure in $X$ & failure in $Z$ & no-info \\ 
 \hline
 Probability & $3\eta^{4}/4$ & $\eta^4/8$ & $\eta^4/8$ & $1-\eta^4$ \\ 
 \hline
 Stabilizers measured & $X_1X_2$, $Z_1Z_2$ & $X_1$, $X_2$ & $Z_1$, $Z_2$ & None\\ 
 \hline
\end{tabular}
\caption{Probability of measuring different stabilizers when performing a linear optical fusion boosted with a Bell pair ($p_{\rm{fail}} = 1/4$) that attempts to measure $X_1X_2$, $Z_1Z_2$ and failure basis randomly chosen between $X_1$, $X_2$ and $Z_1$, $Z_2$. $p_{\rm{loss}}$ is the loss seen by every photon, including boosting photons, and $\eta = 1 - p_{\rm{loss}}$.}
\label{table:fusion_outcomes}
\end{table}

By placing a Hadamard gate before one of the input qubits going into the fusion, we obtain a fusion that measures $X_1Z_2$ and $Z_1X_2$ with both measurements having the same marginal erasure probability of $1 - (1 - p_{\rm{fail}}/2)\eta^4$. This fusion measurement basis is used in the star fusion network in section \ref{sec:example} and the example fusion network in section \ref{sec:fusion-networks}.

\subsection{Encoded fusion erasure probability}
\label{subapp:encoded_fusion_erasure_prob}

In order to increase loss tolerance, the qubits in a resource state are encoded in a four qubit (2,2)-Shor code in section \ref{sec:example}, which is a concatenated of two 2-qubit repetition codes. The (2,2)-Shor code can be oriented in two ways: with code stabilizers $\langle X_1X_2X_3X_4, Z_1Z_3,Z_2Z_4 \rangle$ (X repetition above Z) or $\langle Z_1Z_2Z_3Z_4,X_1X_3,X_2X_4\rangle$ (Z repetition above X).

We perform fusion between two encoded qubits, which we label as $A$ and $B$, by performing pairwise fusions transversally. Considering the case of X repetition above Z, the Shor code stabilizers are $\langle X_{1A}X_{2A}X_{3A}X_{4A}, Z_{1A}Z_{3A}, Z_{2A}Z_{4A}, X_{1B}X_{2B}X_{3B}X_{4B},\\ Z_{1B}Z_{3B}, Z_{2B}Z_{4B} \rangle$. The logical operators for the Shor encoded qubits are $\overline{X_A} = X_{1A}X_{3A}$ and $\overline{Z_A} = Z_{1A}Z_{2A}$, and the same for the B side. The encoded fusion measurements we would like to perform are $\overline{X_A}\overline{X_B}$ and $\overline{Z_A}\overline{Z_B}$. The physical measurements are $X_{iA}X_{iB}, Z_{iA}Z_{iB}, i \in \{1,2,3,4\}$. 

Multiplying with the code stabilizers gives us multiple ways to reconstruct the encoded $XX$ measurement from the physical fusion measurements: 
$$
\overline{X_A}\overline{X_B} = (X_{1A}X_{1B})(X_{3A}X_{3B}) = (X_{2A}X_{2B})(X_{4A}X_{4B})
$$
\noindent where the terms inside the brackets are fusion measurements. There are two ways to measure $\overline{X_A}\overline{X_B}$ and each requires two physical fusion measurements. Therefore, the erasure probability of  $\overline{X_A}\overline{X_B}$ is $[1 - (1-p_0)^2]^2$, where $p_0$ is the physical fusion measurement erasure probability.

Similarly, 

\begin{eqnarray}
\overline{Z_A}\overline{Z_B} &=& (Z_{1A}Z_{1B})(Z_{2A}Z_{2B}) = (Z_{1A}Z_{1B})(Z_{4A}Z_{4B}) \nonumber\\ 
&=& (Z_{2A}Z_{2B})(Z_{3A}Z_{3B}) = (Z_{3A}Z_{3B})(Z_{4A}Z_{4B}) \nonumber
\end{eqnarray}

\noindent i.e. measuring $\overline{Z_A}\overline{Z_B}$ requires one measurement from $\{(Z_{1A}Z_{1B}),  (Z_{3A}Z_{3B})\}$ and one measurement from $\{(Z_{2A}Z_{2B}),  (Z_{4A}Z_{4B})\}$. Therefore, the erasure probability of  $\overline{Z_A}\overline{Z_B}$ is $1 - (1-p_0^2)^2$.

If the orientation of the Shor code is changed, i.e. we have Z repetition above X, the expressions for the encoded $XX$ and $ZZ$ erasure probabilities are flipped. In this paper, we randomly choose the orientation of the Shor code so that the probability of erasure for $\overline{X_A}\overline{X_B}$ and $\overline{Z_A}\overline{Z_B}$ are both

\begin{equation}
p_{\rm{enc}} = \frac{[1 - (1-p_0)^2]^2 + 1 - (1-p_0^2)^2}{2}
\label{eq:penc_fusion_meas_erase_prob}
\end{equation}
When $\eta = 1$ and $p_{\rm{fail}} = 1/4$ as is the case for boosted fusion, $p_0 = 1/8$. This gives us $p_{\rm{enc}} = 0.043$ which is the baseline erasure probability. $p_{\rm{enc}} < p_0$ when $p_0 < 0.5$.

\section{Simulation methods}
\label{sec:simulation_methods}

The hardware-agnostic fusion error model is an i.i.d error model in which every measurement outcome from every fusion is independently erased with probability $p_{\rm{erasure}}$ and suffers a Pauli error (i.e. the measurement outcome is flipped) with probability $p_{\rm{error}}$. Since a measurement can only be incorrect if it is not erased, the probability of the three measurement outcomes are as follows: 
\begin{itemize}
    \item Correct measurement: $(1 - p_{\rm{erasure}}) (1 - p_{\rm{error}})$
    \item Erasure: $p_{\rm{erasure}}$
    \item Incorrect measurement: $p_{\rm{error}} (1 - p_{\rm{erasure}})$
\end{itemize}

\begin{figure}
    \centering
    \includegraphics[width = \columnwidth]{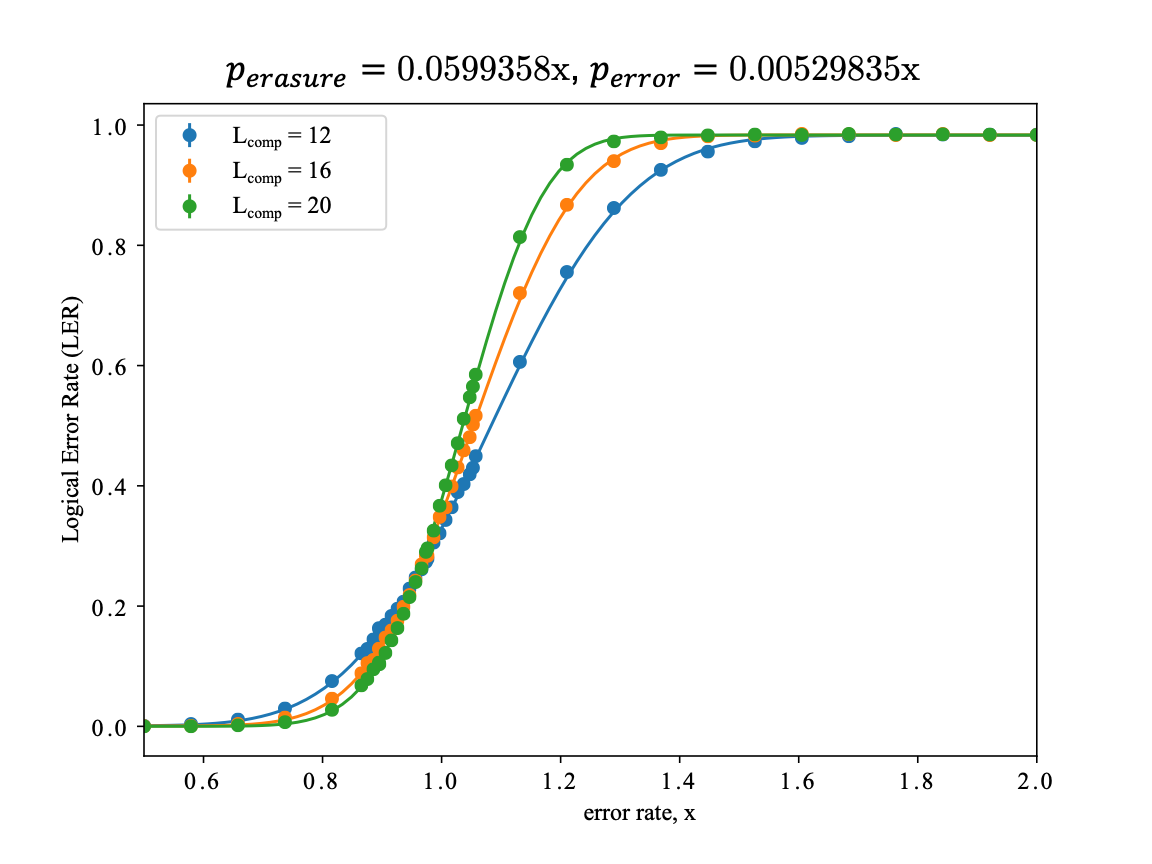}
    \caption{Threshold sweep under the hardware-agnostic fusion error model for the 6-ring fusion network.  We set $p_{\rm{erasure}} = c_{erasure}x$ and $p_{\rm{error}} = c_{error} x$ with $c_{erasure}=0.0599358$ and $c_{error}=0.00529835$, and simulate fusion network blocks of sizes $L_{\mathrm{code}}=${12,16, 20}. Each marker represents one collection of trials for a single set of error parameters. The data is fit to the CDF of the beta distribution. The threshold is the intersection of the curves.}
    \label{fig:example_threshold_sweep}
\end{figure}

To obtain the correctable regions shown in Fig.~\ref{fig:example_phenomenological_phasediagram}, we perform montecarlo simulations of the behavior of the fusion network for different values of $p_{\rm erasure}$ and $p_{\rm error}$. For each combination of error parameters we simulate a 3D block of the fusion network with periodic boundary conditions of size $L_{\mathrm{code}}=${12,16, 20} unit cells in each dimension. To perform a single numerical trial we sample errors and erasures on the edges of the primal and dual syndrome graph according to to the probability distribution above, and compute the resultant syndrome. We then perform decoding using the minimum weight matching decoder on the syndrome graph. We use the Blossom V implementation of graph perfect matching~\cite{kolmogorov2009blossom}. To construct the matching graph we assign a weight 0 to erased edges in the syndrome graph, and all other edges weight 1. We construct a complete matching graph, with a vertex for each odd parity syndrome vertex, and an edge that connects each pair of vertices where the weight of the edge corresponds to the lowest weight path connecting the two vertices on the syndrome graph.  Running the decoder results in a correction, which we compare against the error sample to identify whether a logical error is introduced. We decode primal and dual syndrome graphs separately, and identify the logical error state of both after decoding. We define the overall logical error to be the case that either primal or dual syndrome graph has a logical error in any one of the three periodic dimensions. We repeat this sampling and decoding process at least 15000 times for each combination of error parameters to compute the logical error rate. 
To identify a threshold we fix the two error parameters relative to a single variable $x$: $p_{\rm{erasure}} = c_{erasure}x$ and $p_{\rm{error}} = c_{error} x$ where $c_{erasure}$ are parameters used to vary the ratio of $p_{\rm{erasure}}$ and $p_{\rm{error}}$. We then sweep the value of $x$ and fit to the cumulative distribution function (CDF) of the rescaled and shifted beta distribution to identify a threshold crossing. Fig.~\ref{fig:example_threshold_sweep} shows such a sweep (points) and fit (lines) for the 6-ring fusion network with $c_{erasure}=0.0599358$ and $c_{error}=0.00529835$. The crossing of the different curves here allows us to estimate the threshold for a fixed ratio of $p_{\rm{erasure}}$ and $p_{\rm{error}}$ (equal to $c_{erasure}/c_{error}$), which represents one point in Fig.~\ref{fig:example_phenomenological_phasediagram}. Repeating this for different ratios of $c_{erasure}$ and $c_{error}$, we map out the full curve in Fig.~\ref{fig:example_phenomenological_phasediagram}. 

In the linear optical error model, every photon in the resource state and every boosting photon has a loss probability of $p_{\rm{loss}}$\footnote{When fully accounting for the effect of hardware level errors on photon loss, the loss probabilities of photons used for boosting is likely to be lower than the loss on photons in the resource state. The model we have used is conservative in this respect.}. Every fusion has a failure probability $p_{\rm{fail}}$ which determines the level of boosting required in the fusion and the number of photons involved in each fusion. Following the randomized failure model in Appendix~\ref{app:fusion_error_model}, we find that the erasure probability of both the $XX$ and $ZZ$ measurements for every physical fusion is  $p_0$ (eq.~\ref{eq:p0_fusion_meas_erase_prob}).
If the qubits in the resource state are unencoded, this is the erasure probability of every measurement in the fusion network. If the resource state is encoded in a (2,2)-Shor code as described in Appendix~\ref{subapp:encoded_fusion_erasure_prob}, the erasure probability of every measurement in the fusion network is equal to $p_{\rm{enc}}$ (eq.~\ref{eq:penc_fusion_meas_erase_prob}). Solving the equation $p_{\rm{enc}} = p_{\rm{erasure}}^*$ ($p_0 = p_{\rm{erasure}}^*$) for $p_{\rm{loss}}$, where $p_{\rm{erasure}}^*$, is the erasure threshold of the fusion network, gives us the loss threshold of the fusion network with encoded (unencoded) resource states for a specific value of $p_{\rm{fail}}$. Repeating this for many values of $p_{\rm{fail}}$ gives us the curves in Fig.~\ref{fig:example_lossfail_phasediagram}

\section{Pauli frame}\label{sec:pauli_frame}

During a fusion-based computation we must constantly keep track of the Pauli frame of each logical qubit. After every logical gate, the Pauli frame is updated.

\paragraph{The Pauli Frame}
The {\it Pauli frame} representation of quantum state consists of representing each $n$-qubit quantum state $\psi$ through an $n$-qubit Pauli operator $P$ and a state $\psi'$ such that $\psi = P \psi' P^\dagger$. The Pauli group on $n$ qubits comprises $4^n$ elements, so $P$ can be kept track of classically using $2n$ bits.

The elements in a Pauli orbit of a given stabilizer state, $S$, have the same stabilizer group up to signs. It is this sign that we need to know about during a computation to be able to correctly interpret measurements on logical qubits. 

\paragraph{Where do non-trivial Pauli frames come from.}

Pauli frames emerge during a computation in any architecture for stabilizer-based fault-tolerant quantum computing due to accumulation of physical errors. Decoding syndrome information identifies a correction. It is not necessary to physically implement this correction but rather this can be tracked in the Pauli frame.

In FBQC we must additionally account for the fact that measurement outcomes themselves are intrinsically random. Contrary to the unitary circuit picture, for which (in absence of errors), there is a unique and well defined instantaneous quantum state of the computation $\psi_t$  at any point in time  $t$, this is not the case for FBQC.
As in MBQC, even ideal computations in FBQC proceed with non-determinism in individual measurement outcomes.
In our setting this is due to the fusion measurements which are used to propagate entanglement and correlations through the computation.
Ideally, each Bell fusion can project onto one of four orthogonal entangled states $\{\bra{\Phi^+}, \bra{\Phi^-}, \bra{\Psi^+}, \bra{\Psi^ -}\}$ which are equivalent up to the application of a single qubit Pauli operator. A possible treatment of these fusions is to interpret them as a canonical projection (say, $\bra{\Phi^+}$) and `teleport' an outcome dependent Pauli onto the unmeasured qubits. This is possible assuming that the original state $\ket{\psi}$ being projected has a stabilizer $S$ which can compensate the necessary Pauli on one of the measured qubits.

In fault-tolerant FBQC, the Pauli frame tracks a combination of both of these effects. a) the Pauli correction originating from the intrinsic randomness as well as b) the Pauli correction associated to the most likely Pauli fault equivalence class associated to the extracted syndrome information.

\paragraph{Logical Clifford Gates}

Since logical Pauli operations simply correspond to a change in the Pauli frame, they need not be applied explicitly at the quantum level. Instead they can simply be kept track of using the classical Pauli frame register~\cite{Bertels17}. 
More generally, whenever we wish to implement a quantum gate $U$ on a state $\psi = P \psi' P^\dagger$, we can physically implement $U' \equiv Q U P^\dagger$ for some conveniently chosen Pauli operator $Q$ which will be the new Pauli frame after the gate is implemented.
If $U$ is a Clifford operation, there is no additional cost to this in terms of quantum operations, as the Pauli $Q$ can be chosen to be $Q:=UPU^\dagger$, such that $U':= U$ can be implemented. Pauli product measurements\cite{litinski2019game} are similarly unaffected by the presence of a Pauli frame. The fact that that the same Clifford operations $U$ can be used regardless of the specific Pauli frame is of exceptional importance. It implies that the Pauli frame tracking may lag with respect to the application of the operations themselves, as we discussed in Appendix~\ref{app:decoding}.

The Pauli frame is needed when it comes to logical readout. We must know the Pauli frame in order to correctly interpret logical measurement outcomes.

\begin{figure}[t]
    \centering
    \includegraphics[width=\columnwidth]{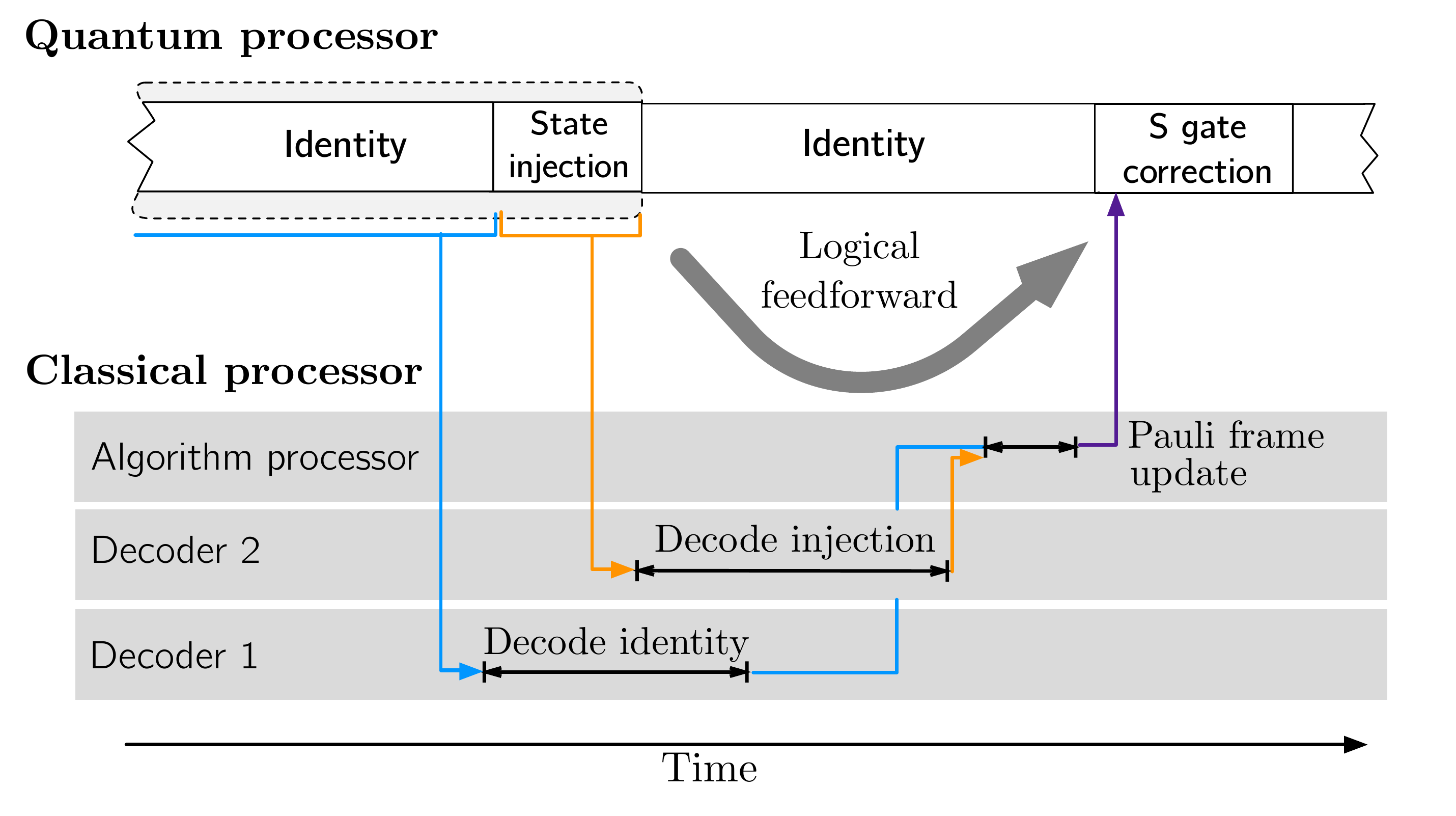}
    \caption{Logical feedfoward in a system where the decoding runtime is roughly twice as long as the logical clock speed. To enable logical feedforward in this situation we can increase the throughput of the decoder by including two decoder processors. In a state injection circuit a logical measurement is made. To interpret this measurement outcome we must combine the decoder outcomes from previous timesteps (indicated by the blue line) with the decoder outcome from the state injection measurement (indicated by the orange line). These solutions can be decoded and then combined to compute the overall Pauli frame, and used to control the correction circuit to complete the state injection. To allow time for this process the logical circuit includes an identity gate which essentially leaves the logical qubit to wait in memory until the information becomes available to implement the next logical gate. It is not necessary to have completed decoding of the second identity operation before implementing the S gate correction. }
    \label{fig:logical_feedforward}
\end{figure}

\section{Decoding}
\label{app:decoding}

The challenges of decoding and handling classical processing and feedforward at the logical level are shared between all models of quantum computation. As we discussed in the main text, FBQC allows a separation of timescales, such that this logical level classical processing can be separated from the timescales of physical measurement reconfiguration. Nevertheless, in the very probable case that decoding and other logical level computation is slower than the physical clock cycle time, the latency of this computation must be accounted for. This can be done through a combination of modifying the logical circuit to allow for decoding latency, and adding additional processors to increase throughput. To illustrate this we can consider the example of the logical feedforward that is needed for a magic state injection circuit. In this case, the state injection involves coupling a target logical qubit with a distilled magic state, making a logical measurement, and then performing a correction circuit dependent on the logical outcome of that measurement. In order to read out the logical measurement we must have completed the decoding up until that point in order to interpret the measurement outcomes. The information needs to become available before the correction circuit is implemented. Figure~\ref{fig:logical_feedforward} illustrates this sequence of events. The crucial timescale here is the latency between receiving the last piece of physical measurement information needed to decode the state injection measurement, and that result being available to configure the circuit to implement the correction circuit. This latency includes the processing time of the decoder, but also the signal transmission times, and the algorithmic level logic about how the decoder outcome will affect the future logical circuit. If this \emph{logical latency} is longer than the physical clock cycle then we can add a buffer region which implements the identity gate until the computation is complete. It is not necessary to have completed decoding of the identity operation before implementing the correction.
If the logical latency is slower than the logical clock speed then in addition to the buffer we need to include additional decoding processors to increase the throughput such that the decoding can be performed at a rate that can `keep up' with the information being produced. The example in Figure~\ref{fig:logical_feedforward} shows the case that the decoding time is roughly twice the logical clock time. To increase the throughput to match the quantum system, we need two decoder processors per logical qubit and we need to pass data to them in an alternating fashion.

\bibliographystyle{unsrt}
\bibliography{fbqc}

\end{document}